\documentclass[12pt,preprint]{aastex}

\slugcomment{Accepted by ApJ}

\def\MO{M_\odot}
\def\LO{L_\odot}

\def\CO{\mathrm{C}^{18}\mathrm{O}}

\def\MSS{{M}_{\mathrm{cluster}}}

\def\MLTE{{M}_{\mathrm{LTE}}}
\def\MVIR{{M}_{\mathrm{VIR}}}

\def\MCLU{{M}_{\mathrm{clump}}}
\def\dv{\Delta{V}}
\def\dvC{\Delta{V}_{\mathrm{clump}}}
\def\RC{{R}_{\mathrm{clump}}}

\shorttitle{Maps of Massive Clumps in the Early Stage of Cluster Formation}
\shortauthors{A.E.Higuchi et al.}

\begin{document}
\title{Maps of Massive Clumps in the Early Stage of Cluster Formation:
Two Modes of Cluster Formation, Coeval or Non-Coeval?}

\author{Aya E. Higuchi\altaffilmark{1,2}, Yasutaka Kurono\altaffilmark{2}, Takahiro Naoi\altaffilmark{2},
Masao Saito\altaffilmark{1,2}, Rainer Mauersberger\altaffilmark{1}, 
\and  Ryohei Kawabe\altaffilmark{1,2}}
\email{ahiguchi@alma.cl}

\altaffiltext{1}{Joint ALMA Observatory, Alonso de C{\'o}rdova 3107, Vitacura, Santiago, Chile}
\altaffiltext{2}{National Astronomical Observatory of Japan 2-21-1 Osawa, Mitaka, Tokyo, 181-8588, Japan}

\begin{abstract}

We present maps of 7 young massive molecular clumps within 5 target regions in the $\CO$ ($J$=1--0) line emission, 
using the Nobeyama 45m telescope. These clumps, which are not associated with clusters, lie at distances between 0.7 to 2.1$\,$kpc.
We find $\CO$ clumps with radii of 0.5--1.7$\,$pc, masses of 470--4200$\,$$\MO$, 
and velocity widths of 1.4--3.3$\,$${\rm{km \: s}^{-1}}$. 
All of the clumps are massive and approximately in virial equilibrium, suggesting they will potentially form clusters. 
Three of our target regions are associated with H {\sc ii} regions 
(``CWHRs" from Clump with H {\sc ii} Regions), and the other two are without H {\sc ii} regions (CWOHRs). 
The $\CO$ clumps can be classified into two morphological types: CWHRs shape a filamentary or shell-like structure, 
CWOHRs are spherical. The two CWOHRs have systematic velocity gradients. 
Using the publicly released $\it{WISE}$ database, Class I and Class I\hspace{-.1em}I protostellar candidates were 
identified within the $\CO$ clumps.
The fraction of the Class I candidates among all YSO candidates (Class I+Class I\hspace{-.1em}I) is $\geq$ 50$\%$ in CWHRs, 
and $\leq$ 50$\,$$\%$ in CWOHRs. We conclude that effects from the H {\sc ii} regions can be seen in 
(1) spatial distributions of the clumps: filamentary or shell-like structure running along the H {\sc ii} regions,
(2) velocity structures of the clumps: large velocity dispersion along shells, and (3) small age spreads of YSOs.
The small spread in age of the YSOs show that the presence of H {\sc ii} regions tend to trigger coeval cluster formation.

\end{abstract}

\section{INTRODUCTION}

Most stars, in particular high-mass stars ($>$$\,$8$\MO$), form as stellar clusters \citep{lad03}. 
Stellar clusters form in dense and massive molecular clumps (size $\sim$$\,$1$\,$pc, mass $\sim$$\,$100--1000$\,$$\MO$, density 
$\sim$ 10$^{3-5}$$\,$$\rm{cm}^{-3}$) with large velocity widths (2--4$\,$km$\,$s$^{-1}$) 
\citep{plu97,rid03, lad03, hig09, hig10}.
Typical median ages of stellar members in these clusters range between 1 and 3 Myr \citep{lad03}, 
although the reason of this spread is unclear.
Since dense gas in the cluster forming clumps is significantly dispersed by the feedback from newly formed stars \citep[e.g.,][]{hig09, hig10},
the initial conditions and details of cluster formation are not understood yet.
Therefore, it is necessary to study molecular clumps in the early stages of cluster formation.
Recently, Infrared Dark Clouds (IRDCs) have been identified as dark extinction features against the bright emission using 
$\it{MSX}$ \citep{ega98, sri05}, and GLIMPSE 8$\micron$ data \citep{per09}.
IRDCs were catalogued by \citet{sim06} and \citet{per09}.
Their physical conditions are similar to those of cluster-forming clumps \citep{rid03, lad03, hig09, hig10}, 
which are the dense regions in molecular clouds.
Most IRDCs are found to be associated with YSO candidates in $\it{Spitzer}$ and $\it{Herschel}$ observations 
\citep{chu06,chu07,deh05,deh10,zav07,zav10}.
Thus, IRDCs are suggested to be precursors of cluster-forming clumps which still preserve initial conditions for the formation of 
stellar clusters \citep{ega98, car98, car00b, rat06, rat10} because the dense gas component is not yet dispersed by the stellar feedback.

Some IRDCs are detected near H {\sc ii} regions \citep[e.g.,][]{deh05, deh10}.
In such cases, cluster formation may be triggered by H {\sc ii} regions.
Recently, ``bubbles", which are surrounding expanding H {\sc ii} regions have been detected from $\it{Spitzer}$ and $\it{Herschel}$ 
\citep{chu06,chu07,deh05,deh10,zav07,zav10}.
The UV radiation from massive stars excite Polycyclic Aromatic Hydrocarbons (PAHs) at the bubble rims and also heat dust grains.
The former can be readily observed at 8$\micron$ by IRAC \citep{chu06,chu07} and the latter at longer 
wavelengths using MIPS on $\it{Spitzer}$ \citep{chu07,wat08}.
Cold dense material has been observed around the borders of the H {\sc ii} regions \citep[e.g.,][]{deh05,deh10,zav07,zav10}.
How the cold dense material can be accumulated by the expanding shells from the H {\sc ii} region has been discussed 
\cite[e.g.,][]{deh05,deh10,zav07,zav10}.
Most of the relevant observations have been conducted with mm dust continuum emission.
Considering dynamical effects such as shock waves and expanding H {\sc ii} region shells should strongly 
influence surrounding molecular gas and change the kinematic structures significantly. 
It is a subject of great interest to investigate velocity structures in molecular gas near the boundary of H {\sc ii} 
regions in a systematic way.

\cite{lop10} presented maps of $\CO$($J$=2--1) toward 48 massive molecular clumps 
and found that about half of their sample sources have velocity gradients of $\sim$ 2 km$\,$s$^{-1}$$\,$pc$^{-1}$. 
\cite{rag12} found similar velocity gradients in NH$_{3}$ lines toward some IRDCs. 
These studies do not focus on the external environments around the clumps.
Understanding the nature and origin of velocity gradients may be crucial in understanding the initial condition of cluster formation.
If a clump is associated with an H {\sc ii} region, 
the velocity structure of the clump will be influenced by shock waves and an expanding shell.
Understanding how clumps are influenced by H {\sc ii} regions require comparison of velocity structures among 
clumps associated with and without H {\sc ii} regions.

In this paper, we present high-resolution observations (effective beam size of $22''$) with the Nobeyama 45m telescope toward seven nearby 
($\leq$ 2.1$\,$kpc) massive clumps (e.g., infrared dark) within five target regions. 
Three of the objects are associated with well-known H {\sc ii} regions (named ``CWHR", i.e., ``Clump with H {\sc ii} Region"), 
the others are without H {\sc ii} region (``CWOHR", i.e., ``Clump without H {\sc ii} Region").
We used the $\CO$($J$=1--0) molecular line emission which traces dense clumps with density around 10$^{3-4}$$\,$$\rm{cm}^{-3}$ regions \citep{hig09}.
$\CO$ is an excellent tracer of the H$_{2}$ column density because in most of cases they are optically thin,
and the relative abundance of CO and its isotopes are in most cases unaffected by the prevailing changing in the gas studied 
(see the discussion in \cite{mau93}). 
We analyzed $\it{WISE}$ archival data available for all target regions, in order to identify YSO candidates which may be associated with clumps, and estimate their relative ages through its Class I fraction in each region.
Here, we investigate the differences of spatial distributions, physical parameters, velocity structures, 
and activity of star formation (e.g., YSO distributions, relative ages) between CWHRs and CWOHRs.

\section{OBSERVATIONS}

\subsection{Source selection}

Our targets are selected from clumps that are expected to be in early stages of cluster 
formation \citep{ega98,beu02a,sri05,rat06}.
Our criteria are that the clumps are massive ($\sim$ 100$\,$$\MO$), which estimated from dust continuum emission, 
and nearby within 2.1$\,$kpc, 
in order to resolve the cloud structures corresponding to the Jeans length (assuming gas temperature of 15 K) 
or finer with the 45m telescope at 110$\,$GHz. 
We finally selected seven young massive clumps (e.g., infrared dark) within 5 target regions, as listed in Table \ref{para}.
The association with H {\sc ii} region or lack of such association is also indicated in Table \ref{para}.

\subsection{Observation with the Nobeyama 45m Telescope}

We carried out $\CO$($J$=1--0; 109.782182 GHz) observations using the 45m telescope 
at the Nobeyama Radio Observatory (NRO) between December 2007 and May 2008.
We used the 25-BEam Array Receiver System (BEARS) for the front end receiver, which has 5$\times$5 beams 
separated by 41$^{\prime\prime}$.1 with respect to each other on the plane of the sky \citep{sun00, yam00}.
At 110 GHz, the FWHM main beam size was 15$''$ and the main beam efficiency was 0.45.
For the back end, we used 25 sets of 1024 channel auto-correlators (AC), each with 32 MHz bandwidth and 
37.8 kHz frequency resolution \citep{sor00}. 
The frequency resolution corresponds to a velocity resolution of 0.10$\,$km $\rm{s}^{-1}$ at 110 GHz.
We applied on-the-fly mapping technique \citep{saw08} and covered areas of 6$^{\prime}$$\times$6$^{\prime}$ 
to 10$^{\prime}$$\times$10$^{\prime}$ which corresponds 
to 3.8$\times$3.8 to 6.1$\times$6.1$\,$pc at a distance of 2.1$\,$kpc. 
During the observations, the system temperatures in the double sideband mode ranged from 400 to 500$\,$K. 
We used emission-free regions near the observed sources as the off positions. 
The standard chopper wheel method was used to calibrate the intensity scale into ${T}^{*}_{\rm{A}}$, 
the antenna temperature corrected for the atmospheric attenuation. 
The accuracy of telescope pointing was checked every 1.5 hr by observing SiO maser sources near 
(within $\sim$ 45$^\circ$) the target regions, and was confirmed to be better than 3$^{\prime\prime}$.
We correct the difference in intensity scales among the 25 beams of the BEARS by using reference data of W51, 
which was obtained in observations with a single sideband filtered SIS receiver S100, and 
the acousto-optical spectrometer (AOS) as the backend.
We used a convolution scheme with a spheroidal function \citep{saw08} to calculate the intensity at 
each grid point of the final map-cube (R.A., Decl., and $v_{\mathrm{LSR}}$) data with a spatial grid 
size of 7$^{\prime\prime}$.5, which is half of the beamsize; the final effective resolution is 
22$^{\prime\prime}$ and the effective integration time at each grid point is about 15 minutes. 
As a result, the typical rms noise level in channel maps for all of the target regions was 0.14 $\pm$ 0.03 K in 0.1 km s$^{-1}$.

We also carried out $\rm{NH}_{3}$(1,1; 23.694495 GHz) and $\rm{NH}_{3}$(2,2; 23.722633 GHz) observations as the backup for bad weather.
At 24 GHz, the FWHM main beam size and the main beam efficiency were 80$''$ and 0.8, respectively. 
For the frontend, we used the 20 GHz HEMT receiver (H20). 
The standard chopper wheel method was used to convert the received intensity into ${T}^{*}_{\mathrm{A}}$. 
The system noise temperature was 100 to 500 K. For the backend, we used the AOSs, which have a velocity resolution of 0.5 km s$^{-1}$ at 24 GHz. 
We observed the $\rm{NH}_{3}$ (1,1) and (2,2) lines simultaneously by using two AOSs each with a bandwidth of 32 MHz. 
The integration time of each point was typically 180 seconds, resulting in a rms noise level in the channel maps of 0.05 K. 
The telescope pointing was checked every 3 hours by observing the SiO maser sources. 
The pointing accuracy was better than 15$''$. 
Data reduction and analysis were made with the NEWSTAR package developed at the Nobeyama Radio Observatory (NRO).

\section{RESULTS}\label{3}

\subsection{Spatial distribution and velocity structure}

\begin{figure}
\epsscale{1}
\plotone{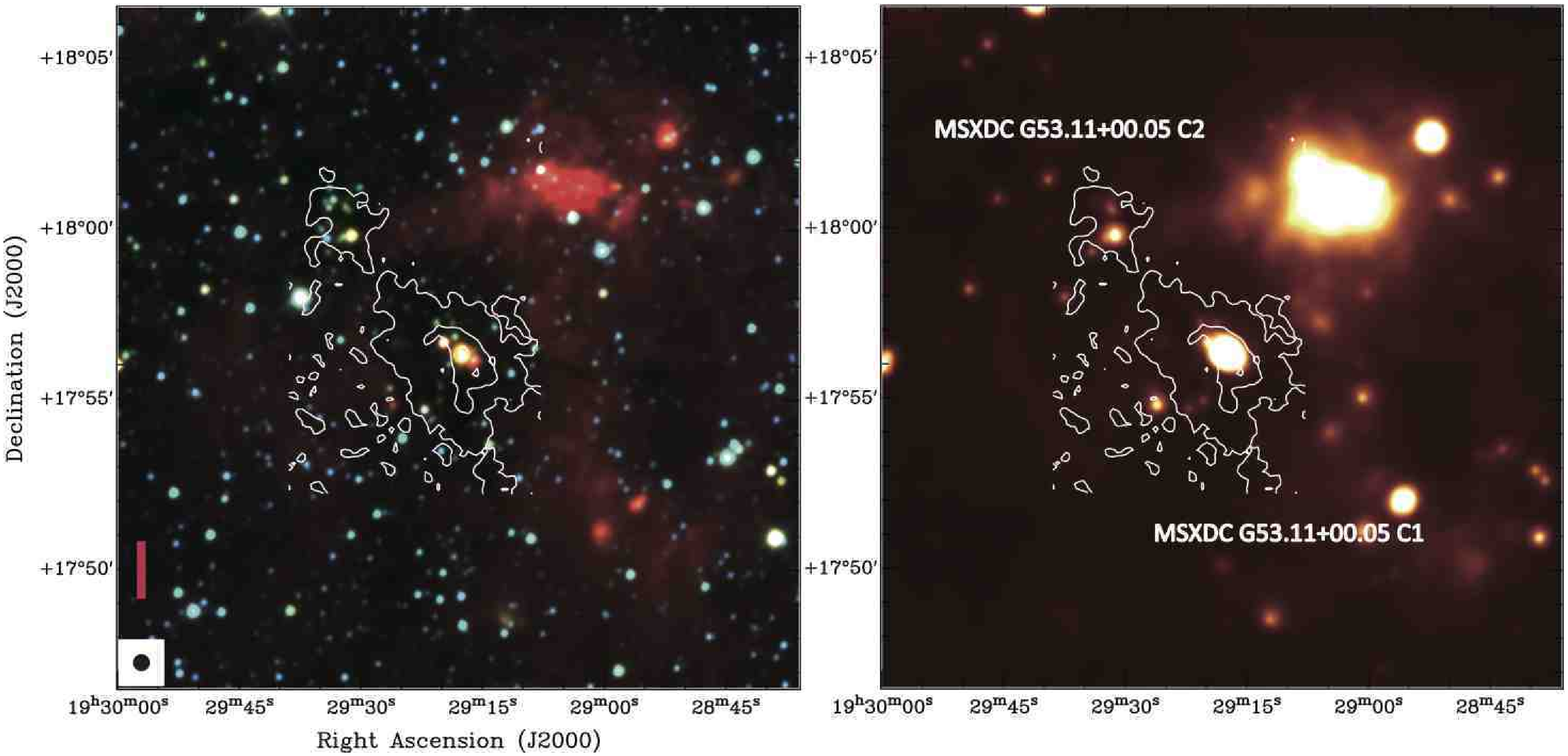}
\caption{[Left] Integrated intensity map of the $\CO$($J$=1--0) emission (contours) superposed on 
the $\it{WISE}$ 3.4 $\micron$, 4.6 $\micron$, and 12 $\micron$  composite color image for MSXDC G 053.11+00.05. 
[Right] Integrated intensity map of the $\CO$($J$=1--0) emission (contours) superposed on 
the $\it{WISE}$ 22 $\micron$ color image.
The contours with the intervals of the 3 $\sigma$ levels start from the 3 $\sigma$ levels for integrated intensity maps 
(1$\sigma$ is 0.53 K km s$^{-1}$).
The pink line shows the 1$\,$pc scale.
The filled circle at the bottom left corner in each panel shows the effective resolution in FWHM of $22^{\prime\prime}$.}
\label{map1}
\end{figure}

\begin{figure}
\epsscale{1}
\plotone{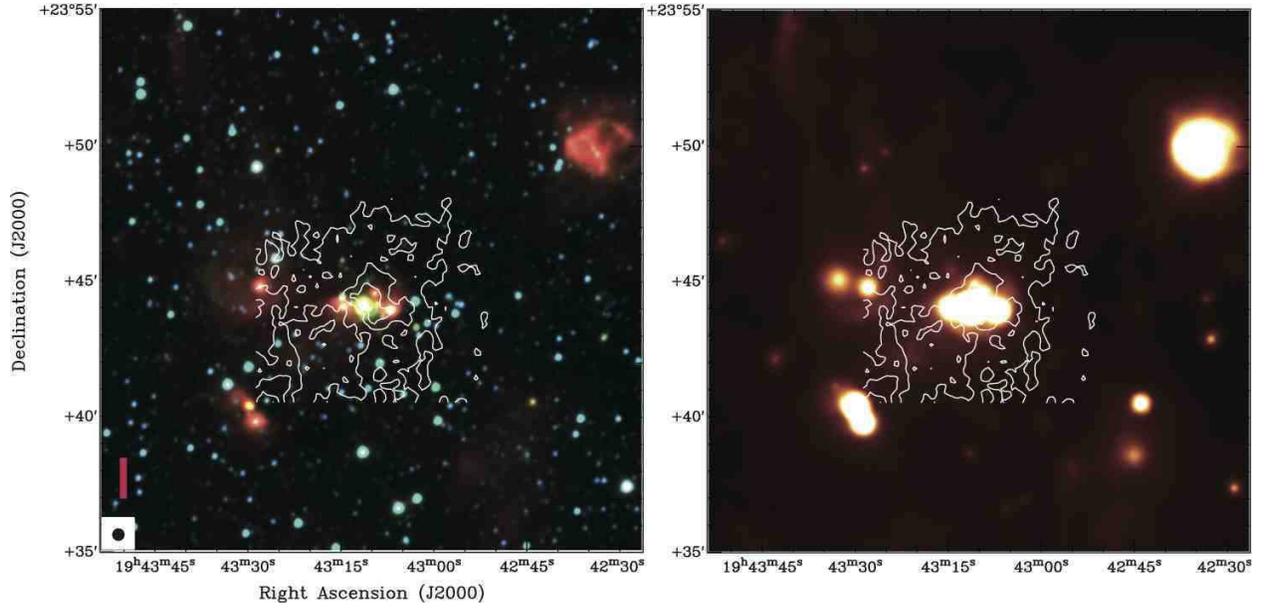}
\caption{Same as Figure \ref{map1} for IRDC 19410+2336. 1$\sigma$ is 0.26 K km s$^{-1}$.}
\label{map2}
\end{figure}

\begin{figure}
\epsscale{1}
\plotone{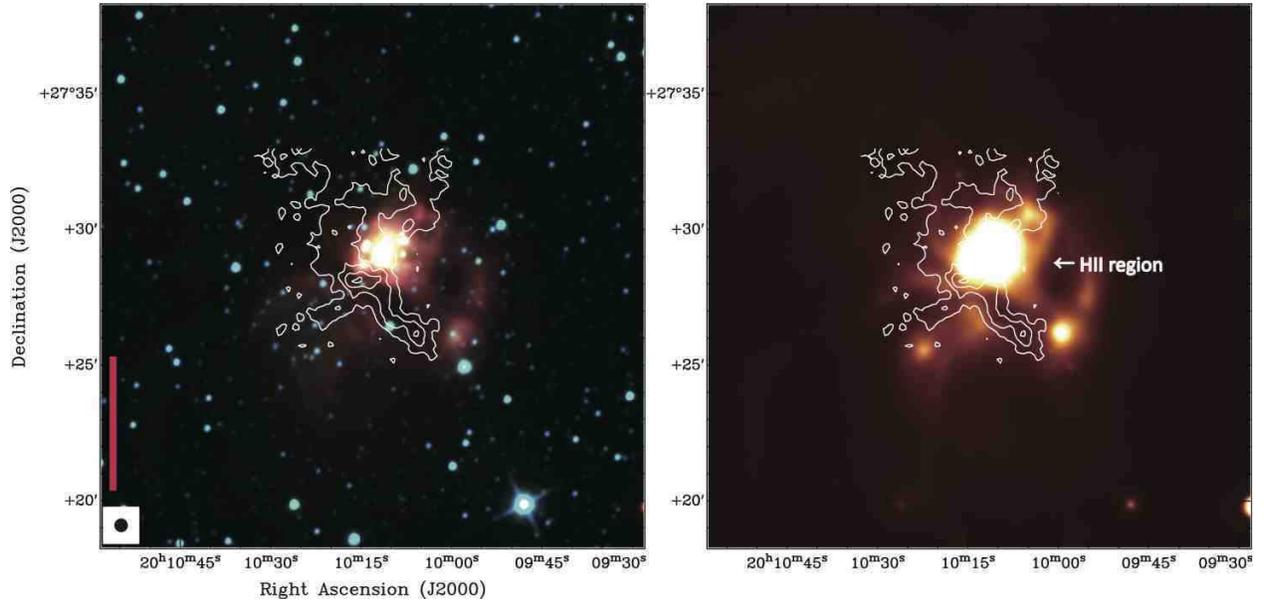}
\caption{Same as Figure \ref{map1} for IRDC 20081+2720. 1$\sigma$ is 0.35 K km s$^{-1}$.}
\label{map3}
\end{figure}

\begin{figure}
\epsscale{1}
\plotone{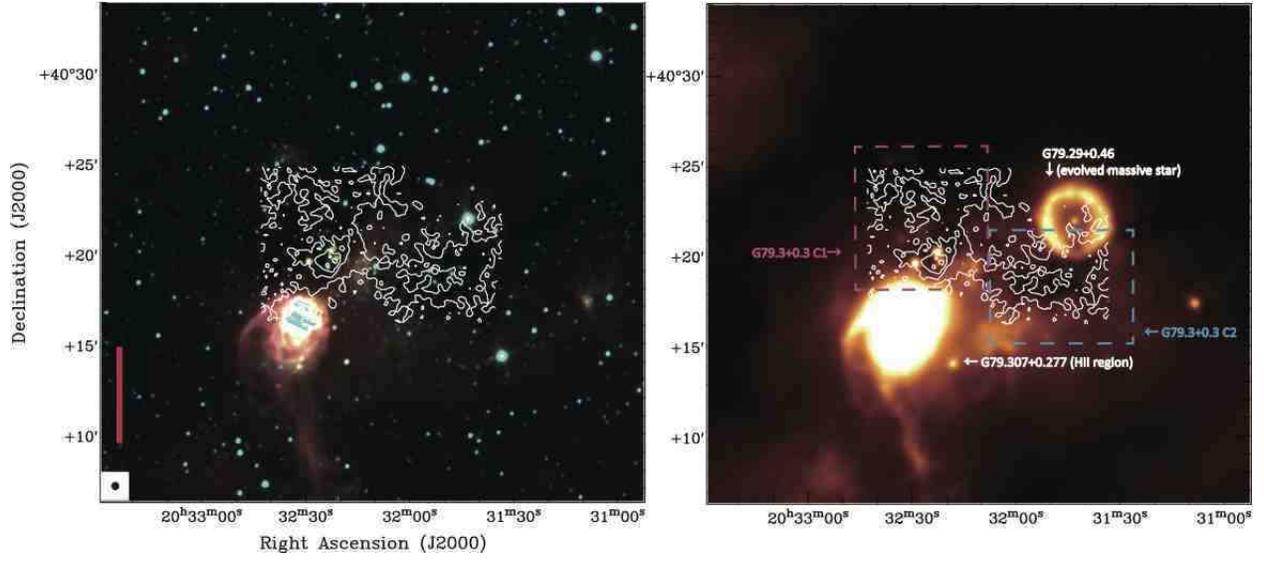}
\caption{Same as Figure \ref{map1} for G 79.3+0.3. 1$\sigma$ is 0.58 K km s$^{-1}$.}
\label{map4}
\end{figure}

Figures \ref{map1} to \ref{map5} show the integrated intensity maps of the $\CO$($J$=1--0) emission 
(contours) superposed on infrared color images in which we used the $\it{WISE}$ 3.4, 4.6, and 12 $\micron$ composite color images, 
and $\it{WISE}$ 22 $\micron$ color images.
We defined the $\CO$ emission enclosed the 3$\,$$\sigma$ contours of the individual map as $\CO$ clump. 
For IRDC 20081+2720, AFGL 333, and G 79.3+0.3, the clumps are associated with well-known H {\sc ii} regions 
(see description of individual sources), while for IRDC 19410+2336 and MSXDC G 053.11+00.05, 
there is no evidence of the existence of H {\sc ii} region
associated with the clumps \footnote{They might be associated with ultra compact H {\sc ii} region within the peak emission of $\CO$}. 
Hereafter, we named the clumps associated with H {\sc ii} regions as ``CWHR" (Clump with H {\sc ii} Region), 
and ``CWOHR" (Clump without H {\sc ii} Region).

From the morphological structure of the $\CO$ clumps, we can classify $\CO$ clumps into two types:
a) filamentary or shell-like structures (IRDC 20081+2720, AFGL 333, and G 79.3+0.3),
b) spherical (IRDC 19410+2336, MSXDC G 053.11+00.05). 
Although the number of samples is small, the CWHRs tend to have the shell-like structures, 
while the CWOHRs have spherical structures from our targets.
In order to investigate the detailed velocity structures of the clumps, 
i.e., presence or absence of velocity gradients, areas of large velocity dispersion,
we produced 1st and 2nd moment maps (Figures \ref{mm1} to \ref{mm5}).
From the 1st moment images, we identify the objects which have a velocity gradient 
($\sim$ 2 km$\,$s$^{-1}$ pc$^{-1}$) within the CWOHRs: IRDC 19410+2336, and MSXDC G 053.11+00.05.
For the CWHRs, the velocity gradient within the clumps can not be well identified.
Furthermore, from the 2nd moment images, we can find that large velocity dispersion ($\sigma$ $\sim$ 0.8 km$\,$s$^{-1}$)
appears along the filamentary structure for AFGL 333 and IRDC 20081+2720, which may indicate the interactions between molecular gas and the expanding shell of the H {\sc ii} region.
For G 79.3+0.3, there are two H {\sc ii} regions (G79.307+0.277: south-east, G79.29+0.46: north-west), 
we can see shell-like structures in the channel maps (0.25--1.2 km s$^{-1}$ in Figure \ref{mapch3}) of the whole of the G 79.3+0.3 clump.

\subsection{Physical properties of $\CO$ clumps}

In this Section, we describe the derivation of the physical properties of the $\CO$ clumps 
(i.e., the area within the 3 $\sigma$ contours).
A summary of the physical properties is given in Table \ref{para}.
To define the clump radius, we estimated the projected area of the clump on the sky plane, $A$ and the observed radius
was derived as $R_{\rm{obs}}$=$(A/\pi)^{1/2}$.
Then we corrected for the spatial resolution as, 
\begin{eqnarray}
\label{radius}
R_{\rm{clump}} \quad=\quad [ R_{\rm{obs}}^2 -
		(\theta_{\rm FWHM}/2)^2]^{1/2} ,
\end{eqnarray}
where $\theta_{\rm{FWHM}}$ of $22''$ the full width half maximum (FWHM) effective beam size.
The velocity width of the clump (one-dimensional FWHM velocity widths), $\dvC$ was derived as follows.
We used the FWHM of the Gaussian fit of spectra averaged over the clumps to derive the virial masses of the clumps (see below).
We derived the $\CO$($J$=1--0) line widths, $\dv_{\mathrm{obs}}$ by fitting the $\CO$($J$=1--0) spectra averaged over the clumps with a Gaussian function. 
We estimated intrinsic velocity widths of the clumps, $\dvC$, by correcting for the 
velocity resolution of the instruments, $\dv_{\mathrm{spec}}$, as $\dvC$ =$\left(\dv_{\mathrm{obs}}^2-\dv_{\mathrm{spec}}^2 \right)^{1/2}$. 
The velocity widths, $\dvC$ range from 1.4 to 3.3$\,$${\rm{km \: s}^{-1}}$, suggesting considerable non-thermal contributions; 
the thermal velocity widths (one-dimensional velocity widths in FWHM) of $\CO$ molecules are 
0.12--0.24$\,$${\rm{km \: s}^{-1}}$ with $T_{\rm{K}}$ = 10--30$\,$K. 
The H$_{2}$ column density of the clumps, ${N_{\rm{H}_{2}}}$ was calculated as,
\begin{eqnarray}
\label{column}
{N_{\rm{H}_{2}}} = 4.7 \times 10^{13} \left(\frac{X_{\CO}}{1.7\times 10^{-7}}\right)^{-1}
	\left(\frac{T_{\rm{ex}}}{\rm{K}}\right)
	\exp \left({\frac{5.27}{T_{\rm{ex}}/\rm{K}}}\right)
	\nonumber \\
	\quad \quad \quad
	\times	
	\left( \frac{\eta}{0.45} \right)^{-1}
	\left( \frac{\tau}{1-\exp({-\tau})} \right)
 	\left( \frac{\int {{T}^{*}_{\rm{A}}} dv}{\rm{K} \: \rm{km \: s}^{-1}} \right)
	\quad [\rm{cm}^{-2}] ,
\end{eqnarray}
where $X_{\CO}$ is the fractional abundance of $\CO$ relative to $\rm{H}_2$, 
$T_{\mathrm{ex}}$ is the excitation temperature of the transition, $\tau$ is the optical depth,
and $\int {T}^{*}_{\mathrm{A}} dv$ is the total integrated intensity of the $\CO$ line emission.
For $X_{\CO}$, we used 1.7 $\times$ $10^{-7}$ by Frerking et al. (1982).
We applied $\tau=0.5$, and ${\tau}/({1-\exp({-\tau})})$ of 1.
Hofner et al. (2000) revealed that most of massive star forming regions are 
optically thin in $\CO$ emission, so that we consider that the assumption is reasonable.
The total mass under the local thermodynamic equilibrium (LTE) condition, $\MCLU$ is calculated as,
\begin{eqnarray}
\label{mass}
\MCLU = 40 \left(\frac{X_{\CO}}{1.7\times 10^{-7}}\right)^{-1}
	\left( \frac{D}{2100 \: \mathrm{pc}} \right)^{2} 
	\left( \frac{\RC}{100 ''} \right)^{2} 
	\nonumber \\
	\quad \quad \quad
	\times
	\left(\frac{T_{\rm{ex}}}{\rm{K}}\right)
	\exp \left({\frac{5.27}{T_{\rm{ex}}/\rm{K}}}\right)
	\left( \frac{\eta}{0.45} \right)^{-1}
 	\left( \frac{\int {{T}^{*}_{\rm{A}}} dv}{\rm{K} \: \rm{km \: s}^{-1}} \right)
	\quad [\MO] ,
\end{eqnarray}
where $D$ is the distance to the object and $\RC$ is the radius of the clump. 
The rotation temperatures, ${T_{\rm{rot}}}$(2,2 ; 1,1) were derived from the NH$_{3}$ data using the same method presented by Ho $\&$ Townes (1983). 
We applied the relation between the kinetic temperature and the rotation temperature, $T_{\rm{K}}$ $\sim$ $T_{\rm{rot}}$ 
as proposed by Danby et al. (1988).
Under the LTE condition, we consider that the excitation temperature 
is comparable to the kinetic temperature ($T_{\rm{ex}}$ $\sim$ $T_{\rm{K}}$).
We used the kinetic temperature derived from NH$_{3}$ data\footnote{Because we could not estimate the excitation temperature for low signal-to-noise ratio of the NH$_{3}$ data (IRDC 19410+2336 and G 79.3+0.3), we adopted 15$\,\rm K$ which is the typical temperature in IRDCs regions.}
as the excitation temperature of the $\CO$ gas.
We also calculated the virial masses of the clumps using ($\MVIR/\MO)=209(\RC/\mathrm{pc})(\dvC/\rm{km \: s}^{-1})^{2}$\citep{ike07}.

We obtained radii of 0.5--1.7$\,$pc, masses of 470--4200$\,$$\MO$, and velocity widths in FWHM of 1.4--3.3$\,$${\rm{km \: s}^{-1}}$ 
(see Table \ref{para}).
We also calculated virial masses, and found that all observed $\CO$ clumps are in 
virial equilibrium within the error in estimate including the uncertainty of abundance ratio and kinetic temperatures.
In order to discuss the kinematics of the clumps, we calculated effective Jeans lengths of 0.12--0.34$\,$pc.
The effective Jeans length, $\lambda_{\rm{J}}$ was calculated as
$\lambda_{\rm J}=0.19\,{\rm pc}\left(T_{\rm K}/10\,{\rm K}\right)^{1/2}\left(n_{\rm H_2}/10^4\, {\rm cm^{-3}}\right)^{-1/2}$, 
where $T_{\rm{K}}$ is the kinetic temperature of the clump and $n_{\rm H_{2}}$ is the density of the clump \citep{sta05}.
The density of the clump, $n_{\rm H_{2}}$ was calculated as $n_{\rm H_{2}}=({3/4\pi})({\MCLU}/\mu{\rm{m_{H_{2}}}}{\RC}^{3})$, 
where $\MCLU$ is LTE mass of the clump, $\mu{\rm{m_{H_{2}}}}$ is mean mass of molecule, and $\RC$ is the radius of the clump.
We also calculated the effective crossing time, 2$\RC$/$\sigma_{\rm{v}}$ ($\sigma_{\rm{v}}$ = $\dvC/\sqrt{8\ln2}$) of (0.9--3)$\times$10$^{6}$$\,$yr,
and the internal kinetic energy, $E_{\rm{kin}}$=(1/2)$\MCLU$${\sigma_{\rm{v}}}^{2}$ 
of (0.17--6.6)$\times$10$^{46}$$\,$erg (see Table \ref{energy}).
Comparing the physical parameters between CWHRs and CWOHRs, there are no significant differences of the physical parameters 
(radii of 0.7--1.7$\,$pc, velocity widths in FWHM of 1.9--3.3$\,$${\rm{km \: s}^{-1}}$, masses of 740--3200$\,$$\MO$ in CWOHRs, and 
0.5--1.4$\,$pc, 1.4--3.0$\,$${\rm{km \: s}^{-1}}$, 470--4200$\,$$\MO$ in CWHRs) and 
kinetic energies ((0.6--6)$\times$10$^{46}$$\,$erg in CWOHRs, (0.2--7)$\times$10$^{46}$$\,$erg in CWHRs)

We calculated the virial ratio using our results, and found that they are gravitationally bound. 
\citet{lad03} suggest that the Star Formation Efficiencies [SFE=$\MSS/(\MSS+\MCLU)$] of the cluster-forming clumps range from approximately 10 to 30$\,$$\%$.
The SFEs of a cluster increases to a maximum value of $\sim$ 30--40$\,$$\%$ due to gas dispersal by the stellar feedback 
\citep{hig09, hig10, lad10}. 
If we assume maximum SFEs of 30$\,$$\%$, cluster masses are expected from 100 to 1000 $\MO$.
\cite{kau10} suggest an empirical mass-size threshold for cluster formation including massive stars. 
Using the mass-size relation; $\MLTE(\RC)$ $>$ 870$\,$$\MO$$\,$$(\RC/\rm{pc})$$^{1.33}$ presented in \citet{kau10}
\footnote{$\RC$ is the effective radius of the clump, $\MLTE(\RC)$ is the mass of the clump}, 
all of the $\CO$ clumps exceed the limit shown in Figure \ref{plot}, 
suggesting that they have a potential to form a cluster including massive stars.

\subsection{YSO candidates in our objects}

In order to investigate the existence of YSO candidates in our target regions, 
we used the information on stellar members in the point source catalogs of the Wide-field Infrared Survey Explorer 
($\it{WISE}$; Wright et al. 2010; Jarrett et al. 2011) for 10$^{\prime}$$\times$10$^{\prime}$ areas.
We adopted the source classification scheme in \cite{koe12}.
They applied the established source classification scheme of $\it{Spitzer}$ IRAC data \citep{gut08, gut09} to the $\it{WISE}$ four bands 
(3.4, 4.6, 12, and 22 $\micron$). 
We selected all sources with photometric uncertainty $<$ 0.2 mag in four bands. 
To reduce the risk that differential extinction will confuse our classification of YSOs, 
we have adopted color-color diagrams of [3.4]--[4.6] vs. [4.6]--[12] and [4.6]--[22] to distinguish between Class I and Class I\hspace{-.1em}I candidates.
Figure \ref{ccd_1} shows source color-color diagrams of [3.4]--[4.6] and [4.6]--[12] made using $\it{WISE}$ 3.4, 4.6, and 12 $\micron$ data.
Class I candidates are identified if their colors match: [3.4]--[4.6] $>$ 1.0 and [4.6]--[12] $>$ 2.0.
Class I\hspace{-.1em}I (T Tauri star candidates) are slightly reddened objects, and were selected by the condition of
[3.4]--[4.6] $>$ 0.25 and [4.6]--[12] $>$ 1.0.
Figure \ref{ccd_2} shows the $\it{WISE}$ 3.4, 4.6, and 22 $\micron$ color-color diagrams.
Class I\hspace{-.1em}I sources are re-classified as [4.6]--[22] $>$ 4.0 \citep{koe12}.
The rest are considered to be diskless sources as in \cite{koe12}.
We selected ``Class I candidates", which are selected from both [3.4]--[4.6] vs. [4.6]--[12] diagram in Figure \ref{ccd_1}, 
and [3.4]--[4.6] vs. [4.6]--[22] diagram in Figure \ref{ccd_2}.

We detected 52 Class I and 28 Class I\hspace{-.1em}I candidates superposed on enclosed 3 $\sigma$ contours of the $\CO$ clumps.
We expect that contamination from non-YSO candidates, such as galaxies (Stern et al. 2005) is small (less than 10 non-YSOs per square degree) 
from the analysis in Allen et al. (2007).
Furthermore, comparing the spatial distributions of YSO candidates with $\CO$ emissions in channel maps, most of YSO candidates are 
associated with $\CO$ emission.
Thus we presume YSO candidates superposed on $\CO$ clumps as a result of star formation within the clumps.
Stellar candidates with intermediate to high masses were identified as the Class I sources with the 
magnitudes lower than 5 in 12 $\micron$ \citep[e.g.,][]{zav07}.
\cite{zav07} identified the brightest sources with $\it{Spitzer}$ 8.0 $\micron$ $\leq$ 6 mag, which corresponds to star that can produce 
Photo Dissociation Regions (PDRs).
Most of extended nebulae are probably local PDRs created by the radiation of the massive stars.
In addition, these are not massive enough to form H {\sc ii} regions, they can heat the surrounding dust and create local PDRs.
Although the wavelength is different from \cite{zav07}, 
the positions of the bright sources with 12 $\micron$ $\leq$ 5 mag corresponds to H {\sc ii} regions or PDRs in our objects.
From these empirical results, 30 $\%$ of detected Class I candidates can be candidates of intermediate to massive stars.
The number density of YSO candidates does not satisfy the definition of ``cluster" in \cite{lad03}
(e.g., 35 or more stars whose space densities exceed 1$\,$$\MO$$\,$pc$^{-3}$).
Consequently, we call these YSO distributions as ``association" in this study.
Note that the AGN and PAH candidates have been removed as in \cite{koe12}, although there is still uncertainty because 
individual candidates have not been identified by the SED. 
Furthermore, in order to check the existence of more evolved YSO candidates, we use the point source catalogs of the 2MASS.
We identified YSO candidates from the 2MASS $J$, $H$, and $K_{\rm{s}}$ data which cover the mapping areas of our dense clump data, 
using the color-color diagrams and methods as in \cite{coh81}, \cite{bes88}, and \cite{car00}.
Figure \ref{ccd} shows $J$--$H$ versus $H$--$K_{\rm{s}}$ color-color diagrams of point sources for the all target regions with 
10$^{\prime}$$\times$10$^{\prime}$ regions. 
The spatial distributions of the YSO candidates are shown in Figures \ref{mm1} to \ref{mm5}.

There is a difference in the spatial distributions of Class I and Class I\hspace{-.1em}I candidates between CWHRs and CWOHRs.
For CWOHRs (IRDC 19410+2336 and MSXDC G 053.11+00.05), the Class I candidates are distributed around 6--9 $\sigma$ contours of $\CO$ emission 
(within $\sim$ 0.5$\,$pc), although it depends on observations.
While, the Class I\hspace{-.1em}I candidates are distributed $\sim$ 2$\,$pc apart from the positions of strong $\CO$ emission, 
which is close to edge of 3 $\sigma$ contours.
For CWHRs (AFGL 333, IRDC 20081+2720), Class I candidates are distributed in the shell-like structure within the clump, 
i.e., they seem to align along the expanding shell from the associated H {\sc ii} region.
For the whole clump of G 79.3+0.3, there seem to be scattering of the distributions of Class I\hspace{-.1em}I candidates compared with 
the distributions of Class I candidates.

The fraction of Class I candidates among YSO candidates, 
${R}_{\rm{Class I}}$=${N}_{\rm{Class I}}$/(${N}_{\rm{Class I}}$+${N}_{\rm{Class I\hspace{-.1em}I}}$) 
was calculated in Table \ref{sep}.
${R}_{\rm{Class I}}$ was found to be higher than 50$\,$$\%$ (50--60$\,$$\%$) in CWHRs, and smaller than 50$\,$$\%$ (10--40$\,$$\%$) in CWOHRs 
(see Figure \ref{histo}).
If we add YSO candidates identified by the 2MASS into calculation of the fraction, this tendency does not change.
In particular, the fraction of Class I candidates among YSO candidates is proportional to the cluster age \citep{eva09, nak11, koe12}.
Comparing the results of 2MASS with {\it{WISE}}, CWOHRs tend to have YSO candidates identified by 2MASS.
This result is consistent with small fraction of Class I candidates among YSO candidates in CWOHRs.
The higher fraction of Class I sources indicates that CWOHRs are relatively younger than CWOHRs. 
Furthermore, this trend implies that the age spread of YSOs in CWHRs is relatively smaller than those of CWOHRs.

\section{DISCUSSION}\label{4}

\subsection{Comparison between CWHR and CWOHR}\label{4-1}

There are three clear signatures that distinguish CWHRs from CWOHRs,
(1) spatial distributions: shell-like or spherical, (2) velocity structures: 
absence or presence of the distinct velocity gradient, large velocity dispersion along shells, 
and (3) age spreads of YSOs: small or large (see Table \ref{score}).
Our results imply that H {\sc ii} regions strongly impact on the spatial distributions, 
velocity structures, and age spreads of YSOs.
However, such an impact is not evident in the other physical proprieties of clumps (size, velocity width, mass, and kinetic energies).

First, the shell-like structure in CWHRs indicates an interaction between the clumps and expansion shell from the H {\sc ii} region.
Comparing the images of dust continuum (also for our targets) and molecular lines (e.g., CO(1--0), $^{13}$CO(1--0)),
the molecular clumps associated with the H {\sc ii} region tend to have shell-like or filamentary structures \citep{deh05,deh10,bra11}.
These morphological structures may have been shaped by an H {\sc ii} region.
Filaments will be unstable due to the interactions and will be easy to fragment \citep[e.g.,][]{tei06,tei07,pie11}.
\cite{fuk00} presented that there are two factors that can trigger star formation by an H {\sc ii} region; 
dynamical compression due to the H {\sc ii} region and change in the self-gravity.
In a numerical study of star formation, \cite{hos06} suggest that an expanding H {\sc ii} region 
is able to trigger star formation within the ambient molecular clouds traced by CO(1--0), $^{13}$CO(1--0) lines.
We discuss this in more detail in Sec.\ref{4-2-1}.

Second, regarding the difference of velocity structures, 
shock waves and the expanding shell of an H {\sc ii} region should strongly influence surrounding
molecular gas and change significantly the kinematic structure of the clumps.
\cite{bra11} present maps with various molecular lines around the H {\sc ii} region, Sh 2-217.
They see no obvious velocity gradients.
Velocity structures in CWHRs has not been discussed previously because most of previous surveys of dense material around H {\sc ii} 
regions have been conducted with mm-continuum.
This is discussed in more detail in Sec.\ref{4-2-2}.

Third, the value of ${R}_{\rm{Class I}}$ is considered to be a clear 
indication of the relative evolutionary status of the YSO association \citep{eva09, nak11, koe12}.
\cite{koe12} present a plot of the ratio, ${R}_{\rm{Class I}}$ for the cluster-forming regions as a function of the central cluster age.
Using their correlation, the ages of our stellar associations obtained by {\it{WISE}} are less than 1$\,$Myr.
In particular, the ages of the associations are estimated to be less than 0.5$\,$Myr if the 
value of ${R}_{\rm{Class I}}$ is higher than 50$\,$$\%$.
The higher fraction of Class I sources indicates that CWOHRs are relatively younger than CWOHRs. 
Furthermore, this trend implies that the age spread of YSOs in CWHRs is relatively smaller than those of CWOHRs.
Comparing the spatial distributions of YSO candidates with the $\CO$ emissions, 
Class I candidates are distributed uniformly in the shell-like structure in CWHRs.
On the other hand, the Class I candidates are distributed with the peak emission in $\CO$, 
and Class I\hspace{-.1em}I candidates are distributed outward of the Class I candidates in CWOHRs.
\cite{zav07} discussed the triggering mechanism of an H {\sc ii} region, RCW 120 (d $\sim$ 1.3$\,$kpc), 
and identified YSO candidates using {\it{Spitzer}} data.
The spatial distribution of YSO candidates within ``condensation 1" around RCW 120, 
whose distance are $\sim$ 0.8$\,$pc from the exciting stars, are similar to those of G 79.3+0.3, IRDC 20081+2720, and AFGL 333.
Class I candidates are uniformly distributed and aligned on the shell structure \citep{zav07, pug09}.
From the spatial distribution of the YSO candidates toward RCW 120 (Figure 5 in \cite{zav07}), 
we calculated the value of $R_{\rm{Class I}}$ and found it to be $\sim$ 50$\,$$\%$, which is consistent with our results in CWHRs.

\subsection{Triggered cluster formation on the periphery of H {\sc ii} regions}\label{4-2}
\subsubsection{Triggering mechanism by H {\sc ii} regions}\label{4-2-1}

Star formation may be triggered by the expanding shell motion of an H {\sc ii} region for AFGL 333, 
IRDC 20081+2720, and the whole clump of G 79.3+0.3.
Dynamical effects of H {\sc ii} regions can be seen in the spatial distributions, 
velocity structures, and small age spreads of YSOs.
Here, we discuss a possible triggering mechanism, which leads these different signatures.

There are mainly two models proposed for triggered star formation by H {\sc ii} regions, 
radiation-driven implosion \citep{lef94,kes03} and collect-and-collapse model \citep{elm77,zav06,deh10}.
In the collect-and-collapse model, swept-up molecular gas is accumulated to form dense gas in the shock front 
by the supersonic expansion of an H {\sc ii} region. 
The shell is mainly occupied by dense molecular gas whose mass increases with evolution.
\cite{zav07} discussed the dynamical expansion of RCW 120 using the model of \cite{hos06}. 
Figure 14 shows the gas shell swept up by the shock front, with densities in the range of 10$^{4-5}$$\,$cm$^{-3}$.
This layer may become unstable and fragment, which make dense core (size $\sim$ 0.1$\,$pc, density $\sim$ 10$^{4-5}$$\,$cm$^{-3}$; \cite{elm77, whi94}). 
This process tends to sweep up amounts of gas and allows to form massive dense clumps \citep{bra11}, 
which is considered to form clusters including massive stars.
On the other hand, radiation-driven implosion models are predicted to form only low- to intermediate stars \citep{deh10, bra11}. 

Our clumps are massive enough to form a cluster including massive stars \citep{kau10}.
In addition, CWHRs have a few candidates of intermediate to massive stars are distributed around the H {\sc ii} region 
in G 79.3+0.3 and AFGL 333.
From those results, we suggest that the collect-and-collapse model may be more adequate than the radiation-driven implosion.

\subsubsection{Kinematic energy injection from H {\sc ii} regions}\label{4-2-2}

Let us assume that there is one 20$\,$$\MO$ star, which is located at 0.5$\,$pc 
(the separations between the positions of H {\sc ii} region and the strong $\CO$ emission peak are 0.3 to 1$\,$pc) 
apart from the strong $\CO$ emission peak of the clumps. 
We estimate the internal kinetic energy, ${E}_{\rm{kin}}$ of the individual molecular clumps to be $\sim$ 10$^{45-46}$$\,$erg (see Table \ref{energy}).
We consider the kinetic energy transferred to the clump through the stellar wind. 
Using the values of \cite{abb82}, we estimated the kinetic energy supply rate by the stellar wind from a 20$\,$$\MO$ star to be $\sim$ 10$^{35}$ erg$\,$s$^{-1}$.
Here, we assume that the stellar wind is isotropically driven to the molecular clump.
The kinetic energy of stellar wind during the crossing time ($E_{\rm{SW}}$ in Niwa et al. 2009) 
is estimated to be $\sim$ 10$^{48}$ erg, which is larger than internal kinetic energies of the clumps.
Therefore our estimation suggests that H {\sc ii} regions have potential to change the kinematics of the clumps 
located on the border of the H {\sc ii} region.
We can present that the energy input from the H {\sc ii} regions through the shocks of stellar
wind are sufficient to drive the internal kinetic energy of the clump.

\subsection{Modes of cluster formation: coeval or non-coeval}\label{4-3}

Triggered star formation by an H {\sc ii} region can explain small age spreads of YSOs.
According to \cite{zav07}, the dynamical expansion of an H {\sc ii} region 
(excited by a star of $\sim$ 20 $\MO$) reaches a size of 1$\,$pc in 0.2$\,$Myr.
An H {\sc ii} region will accumulate dense gas and derive the formation of filaments on the border of the H {\sc ii} region.
Dynamical interactions between an H {\sc ii} region and a filament will occur.
This filament will become gravitationally unstable due to the interaction. 
Once dense filament forms, fragmentation will occur along the filament to form dense cores, 
whose typical separation are the order of the Jeans length \citep{tei06,tei07,pie11}.
This mechanism will coevally form dense cores along the filament.
These cores will immediately start collapsing in the free-fall time \citep{ike07}, which explains the small age spread of YSOs in CWHRs.
In fact, fragment components can be seen in the channel maps around the H {\sc ii} region in Figure \ref{mapch5} to \ref{mapch4}.
YSO candidates with the separations of the order of Jeans length are distributed along the filament in observed CWHRs.

For CWOHRs, the clumps have spherical structure.
They are more stable than filamentary one \citep[e.g.,][]{inu92}.
Nevertheless, a spherical clump will eventually fragment, which initiates cluster formation \citep[e.g.,][]{har02,bon03}.
After fragmentation, the local condensations will become distinct peaks, i.e., dense cores, where stellar clusters will be eventually born. 
\cite{hig09, hig10} classified 14 cluster-forming clumps into three evolutionary stages and found that spherical clumps in the early 
stage tend to have a single peak. This result may imply that the fragmentation rate in a spherical clump can be lower than in a filament.
During the cluster formation, the cores within the clumps will form stars in a wide mass range and dense gas is dispersed 
by their stellar activities \citep{hig09, hig10}.
Once clusters are formed, young stars, particularly most massive ones will disperse the surrounding gas with evolution making cavities around them \citep{fue98,fue02}.
Stellar feedback also forms dense cores \citep[e.g.,][]{kne00}, the structures of the evolved clumps are expected to be less condensed than those of the younger clumps. 
The differences of the fragmentation rate between CWHRs and CWOHRs will explain the difference of the age spreads of YSOs.

\cite{pom09} observed swept up dense gas surrounding H {\sc ii} region, RCW 82, and several Class I candidates along the expanding shell.
\cite{pug09} investigated the spread of ages in Sh-284 with $\it{Spitzer}$/IRAC, and presented that despite the different scales ($\sim$ 20$\,$pc) observed in its multiple-bubble morphology, the spread of ages among the powering high-mass clusters is $\sim$ 3$\,$Myr.
\cite{bik12} presented the evidence of an age spread of at least 2--3$\,$Myr in W3 Main.
Those scales of H {\sc ii} regions are a magnitude larger than those of our objects.
If we use the age spread-linear size relationship described by \cite{elm00}, 
the age spread is roughly estimated to $\sim$ 0.5$\,$Myr (i.e., consistent with free-fall time) in our observed scales.
This value is consistent with the ages of associations derived from relation between the ratio of Class I candidate and cluster age in \cite{koe12}.
Lifetimes of the low to intermediate stars (Class I and Class I\hspace{-.1em}I candidates) are considered to be $\sim$ 10$\,$Myr \citep{sta05}.
Thus, an age spread of 0.5$\,$Myr is significantly small to speak of ``coeval" cluster formation.

We propose that there are two modes of cluster formation; one is coeval cluster formation, the other is non-coeval one.
This difference is related to their environments (existence or non-existence of the H {\sc ii} region) of the associated clumps.
H {\sc ii} regions play three roles:
(1) they accumulate the dense gas by an expanding shell, (2) they shape the dense component into a filament, and (3) they inject energy.
Our study can reveal differences between the mode of cluster formation as we focused on clumps in the earliest stages of cluster formation.
Interestingly, both CWHRs and CWOHRs have potential to form a cluster containing massive stars.
In particular, CWOHRs contain candidates of intermediate to massive stars, although their fraction of Class I\hspace{-.1em}I candidates is relatively higher.
If the differences of cluster-forming modes depend on the differences of the environments of the clumps,
some unknown issues of cluster formation (e.g., IMF, star formation efficiency, age spread of the cluster members)
can be clarified by carrying out such an investigation toward the clumps in the early stages of the cluster formation (e.g.,$\,$IRDCs).
Although more statistical studies should be needed, these studies will lead the understanding the fundamental 
topic of the cluster formation.

\section{CONCLUSIONS}

We have mapped 7 young massive molecular clumps within 5 target regions 
in the $\CO$ line emission with the Nobeyama 45m radio telescope.
Our aim of this study is to understand the dynamical impact of H {\sc ii} regions on the clumps.
We found different signatures of the clumps with H {\sc ii} region as compared to clumps 
or without one. 
Our results and conclusions are summarized as follows:

\begin{enumerate}

\item We made the $\CO$ maps with a size of $\sim$ 6$^{\prime}$$\times$6$^{\prime}$ to 10$^{\prime}$$\times$10$^{\prime}$ areas of 7 nearby 
($D$ $\leq$ 2.1$\,$kpc) young massive clumps within 5 targets.
As a result, all the clumps are detected in $\CO$ lines, 
whose radii, masses, and velocity widths are 0.5--1.7$\,$pc, 470--4200$\,$$\MO$, and 1.4--3.3$\,$km s$^{-1}$, respectively. 
They are similar to those of the cluster-forming clumps.

\item 
All the clumps are likely to be in virial equilibrium, 
and exceed the massive star formation limit; mass-size relation presented in \citet{kau10}, 
suggesting that they have a potential to form a cluster including massive stars.

\item
In order to investigate the existence of YSO candidates in our target regions, 
we identified stellar members using the point source catalogs of the {\it{WISE}}. 
There are 52 Class I and 28 Class I\hspace{-.1em}I candidates within the all $\CO$ clumps. 
30 $\%$ of the detected Class I candidates may be intermediate to massive stars.
The fraction of Class I to YSO candidates: 
${R}_{\rm{Class I}}$=${N}_{\rm{Class I}}$/(${N}_{\rm{Class I}}$+${N}_{\rm{Class I\hspace{-.1em}I}}$) is presented in Table \ref{sep}.
It is $\geq$ 50$\%$ (50--60 $\%$) in CWHRs, and $\leq$ 50$\%$ (10--40 $\%$) in CWOHRs.

\item 
There are three clear signatures that distinguish CWHR and CWOHR,
(1) spatial distributions: shell-like or spherical, (2) velocity structures: 
absence or presence of the distinct velocity gradient, large velocity dispersion along shells, 
and (3) the age spread of YSOs: small or large.

\item
An H {\sc ii} region can play three roles:
(1) accumulate the dense gas by expanding shell, (2) shape the dense component into a filament, and (3) inject energy.
The expanding shell can accumulate dense gas around the border of H {\sc ii} region and make 
the dense gas filament. 
Interaction between H {\sc ii} regions and filaments will tend to form dense core coevally.
Once dense cores form, these cores will immediately start collapsing in a free-fall time, 
which can explain the small age spread of YSOs.

\end{enumerate}

\bigskip
We thank the anonymous referee for constructive comments that helped to improve this manuscript. 
We are grateful to the staff of the Nobeyama Radio Observatory 
(NRO)\footnote{Nobeyama Radio Observatory is a branch of the National Astronomical Observatory of Japan, 
National Institutes of Natural Sciences.} for both operating the Nobeyama 45m telescope and helping us with the data reduction.
We also thank S.Komugi for useful discussion.

\appendix
\section{Individual sources}\label{source}

\subsection{MSXDC G 053.11+00.05}

From the $\it{WISE}$ composite image (Figure \ref{map1}), 
we can see two main YSO associations in MSXDC G 053.11+00.05 region located on north and south (also see Figure \ref{mm1}). 
The channel maps in Figure \ref{mapch1} clearly show that strong $\CO$ line emission is associated with these YSO candidates 
in different velocity ranges, for the south one which is brighter is in a range of 20.8 km s$^{-1}$ to 22.8 km s$^{-1}$, 
for the other in north is in a range of 20.8 km s$^{-1}$ to 24.2 km s$^{-1}$. 
The $\CO$ clump seems to have a velocity gradient along the north-south direction. 
We consider that these emission components correspond to parental clumps of these YSO associations (see Figure \ref{mm1}), 
which are labeled as C1 for the south one and C2 (corresponds to MSXDC G 053.25+00.04 in \cite{rat06}) for the north one. 
For clump C1, compact dust condensation is detected by \citet{rat06} in 1.2 mm continuum observations.
MSXDC G 053.11+00.05 C1 have region of large velocity dispersion where YSO candidates are distributed,
while MSXDC G 053.11+00.05 C2 have the large velocity dispersion where YSO candidates are not distributed.

\subsection{IRDC 19410+2336}

From the images of Figure \ref{map2}, we can see a YSO association in the $\it{WISE}$ composite image around 
IRDC 19410+2336 central region, corresponding to IRAS 19410+2336 \citep{mar08}.
The channel maps in Figure \ref{mapch2} clearly show that strong $\CO$ line emission is associated with YSO candidates 
in a velocity range from 21.2 km$\,$s$^{-1}$ to 23.2 km$\,$s$^{-1}$, and the $\CO$ clump seems 
to have a velocity gradient in the range from 20.8 km$\,$s$^{-1}$ to 23.8 km$\,$s$^{-1}$ in the direction from north-east to south-west. 
The multiple CO($J$=2--1) outflows (blue; 5--18 km$\,$s$^{-1}$, red; 26--47 km$\,$s$^{-1}$) have been detected in the direction 
from north-east to south-west by \citet{beu03}, which implies that IRDC 19410+2336 is a young active star forming region.

\subsection{IRDC 20081+2720}

There is bright infrared emission in the $\it{WISE}$ composite image (Figure \ref{map3}), which corresponds to an H {\sc ii} region \citep{beu02a}.
The H {\sc ii} region is located $\sim$ 0.3$\,$pc away from the strong $\CO$ line emission.
In a velocity range from 5.2 km$\,$s$^{-1}$ to 5.8 km$\,$s$^{-1}$, 
we can see the $\CO$ emission have shell-like structure, which may be shaped by the H {\sc ii} region, i.e., from the 
interaction between dense gas and expanding shell.
The YSO candidates are distributed within the shell-like structures of $\CO$ emission around the H {\sc ii} region.

\subsection{G 79.3+0.3}

There is a bright H {\sc ii} region (Figure \ref{map4}), 
DR 15 (G79.307+0.277), which lies behind and slightly south ($\sim$ 0.6$\,$pc) of the G 79.3+0.3 region \citep{red03}.
DR 15 is formed by one or two B0 zero-age main-sequence stars, observed as the far-infrared source 
(Odenwald et al. 1990), having a total luminosity of $\sim$ 3 $\times$ 10$^{6}$$\,$$\LO$ (Oka et al. 2001).
A ring shaped nebula in 22 $\micron$ emission corresponds to a photo-dissociation region, G79.29+0.46, 
which is the outskirts of an evolved massive star \citep{riz08}.
From the morphological distributions between $\CO$ and 22 $\micron$ emission, in a velocity range from 
$-$0.25 km$\,$s$^{-1}$ to 1.2 km$\,$s$^{-1}$, we can see the $\CO$ emission have an shell-like structure in the north area, 
which is an possibly compressed by G79.29+0.46.
The YSO candidates are distributed in a shell-like fashion around the H {\sc ii} region (DR 15).
The channel maps in Figure \ref{mapch3} clearly show that strong $\CO$ line emission is associated with YSO candidates 
for the east one in a range of 0.25 km$\,$s$^{-1}$ to 2.8 km$\,$s$^{-1}$.
For the other condensations in west, it is still dark in the 22 $\micron$ image.
In G 79.3+0.3 region, the global trend of velocity gradients can not be well defined.
The clump of the association for the east one are labeled as C1 (correspond to P1 in \cite{red03}).
For the west one labeled as C2 (correspond to P3 in \cite{red03}), the $\CO$ clump is 
associated with an infrared dark area.

\subsection{AFGL 333}

\begin{figure}
\epsscale{1}
\plotone{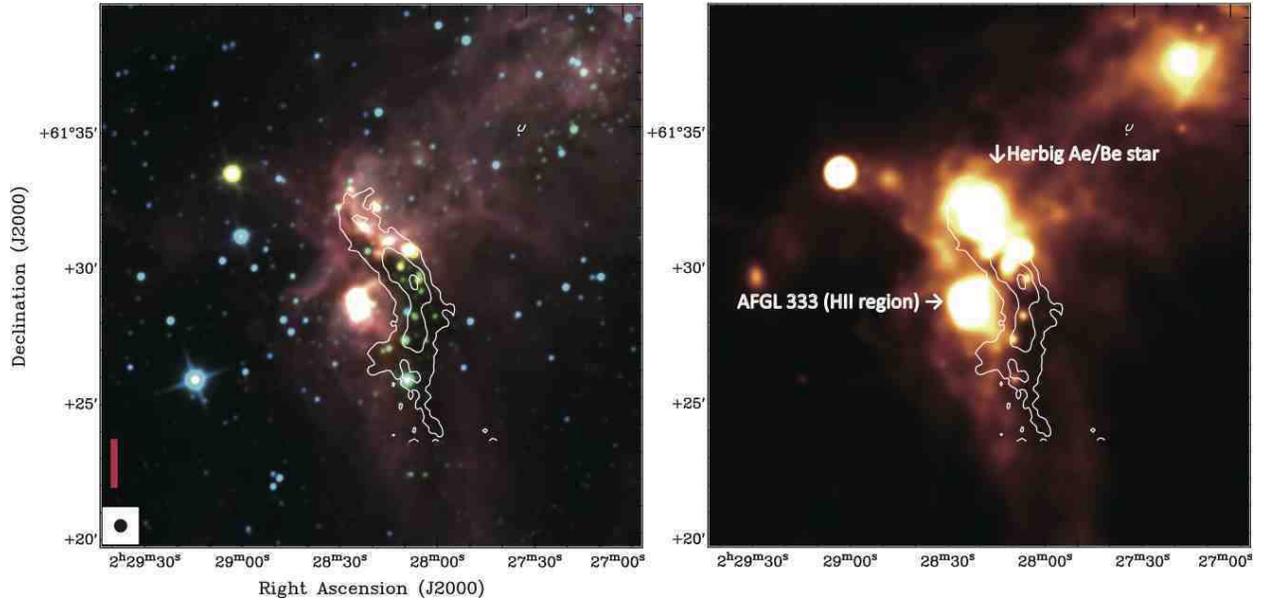}
\caption{Same as Figure \ref{map1} for AFGL 333. 1$\sigma$ is 0.62 K km s$^{-1}$.}
\label{map5}
\end{figure}

\begin{figure}
\epsscale{0.8}
\plotone{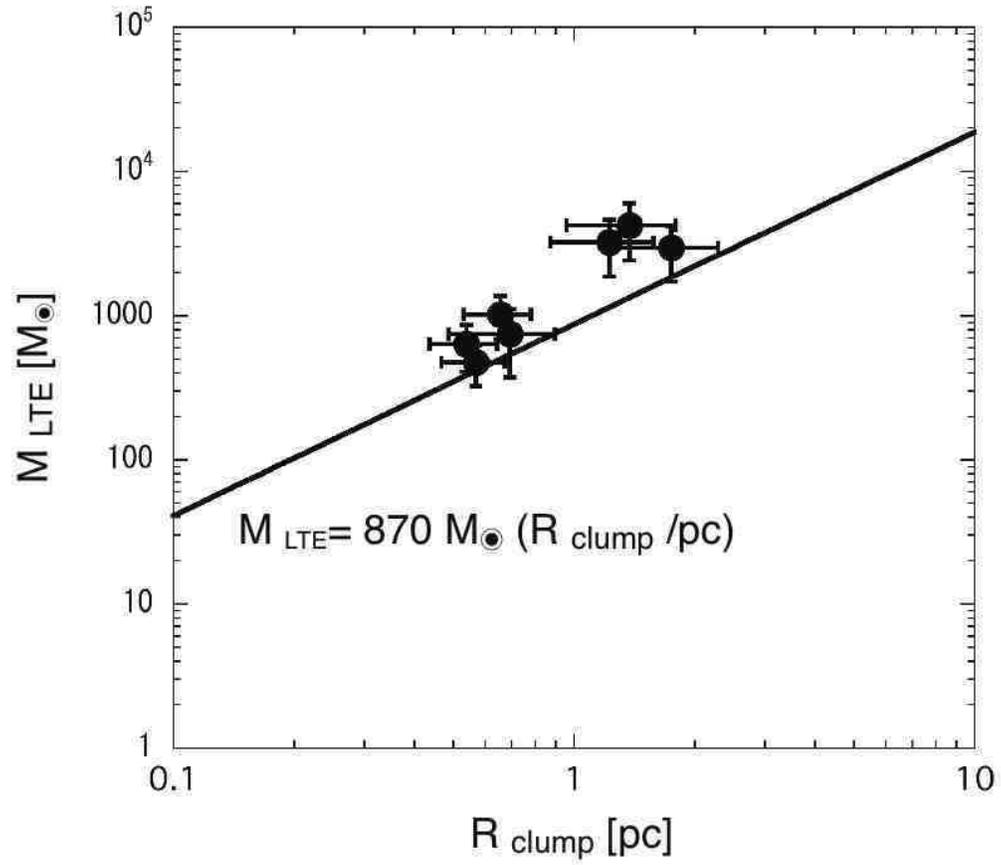}
\caption{Plot of radius vs. mass and the mass-radius relation (solid line) derived in \citet{kau10}.
The mass-radius relation up to which clumps form massive stars.}
\label{plot}
\end{figure}

\begin{figure}
\epsscale{1}
\plotone{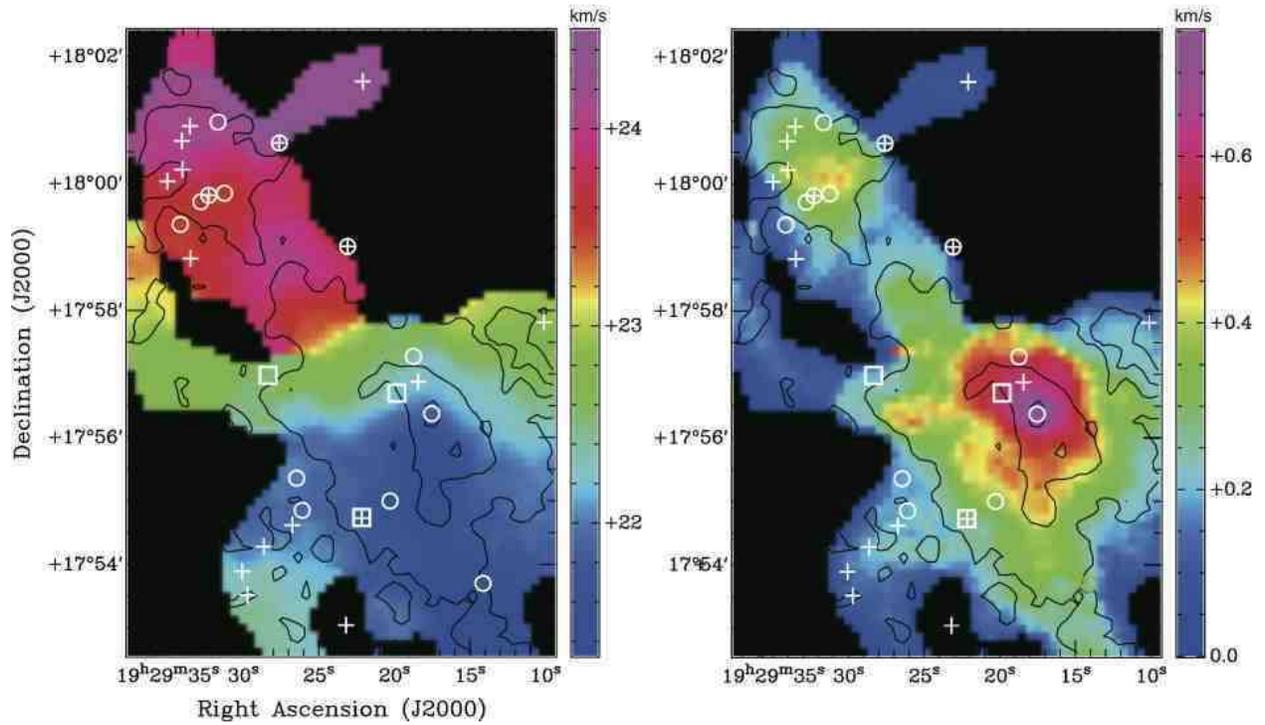}
\caption{(Left) The integrated intensity map of the $\CO$($J$=1--0) emission (contours) superposed on the 1st moment map (color).
(Right) The integrated intensity map of the $\CO$($J$=1--0) superposed on the 2nd moment map (color) for MSXDC G 053.11+00.05.
Small circles show Class I candidates, cross marks show Class I\hspace{-.1em}I candidates, 
and large squares show YSO candidates identified by 2MASS.}
\label{mm1}
\end{figure}

\begin{figure}
\epsscale{1}
\plotone{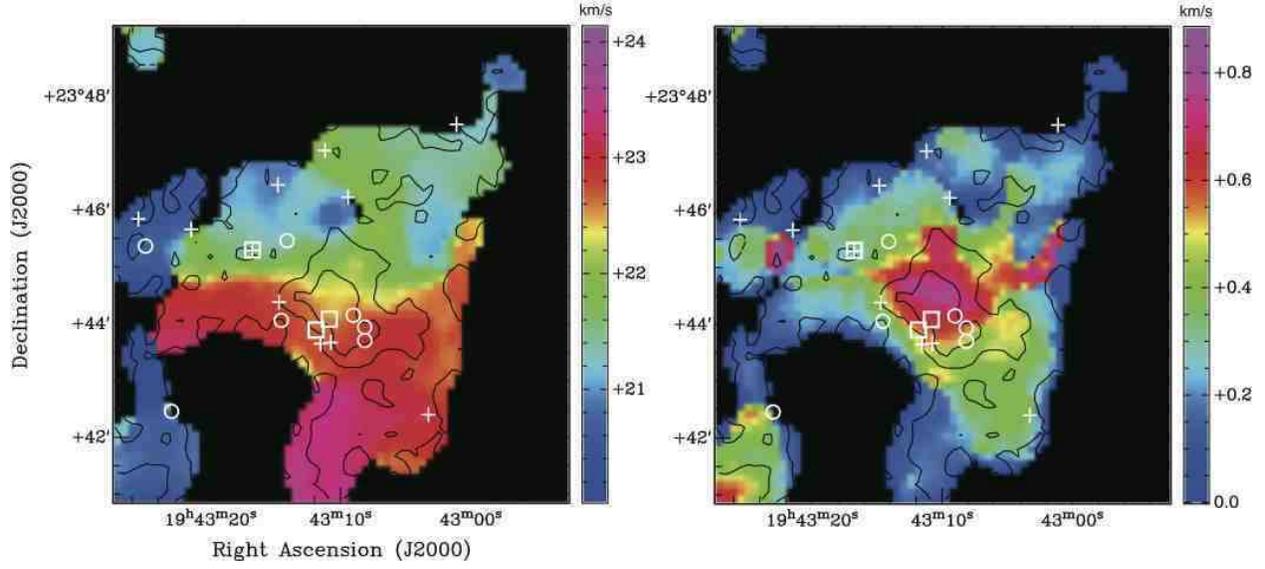}
\caption{Same as Figure \ref{mm1} for IRDC 19410+2336.}
\label{mm2}
\end{figure}

\begin{figure}
\epsscale{1}
\plotone{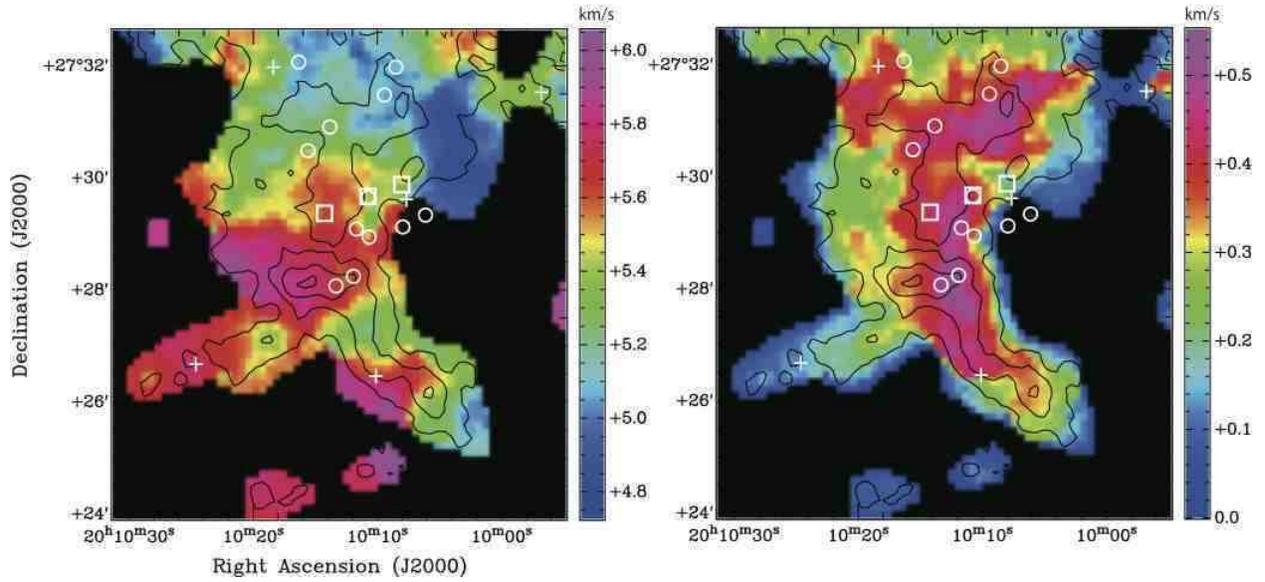}
\caption{Same as Figure \ref{mm1} for IRDC 20081+2720.}
\label{mm3}
\end{figure}

\begin{figure}
\epsscale{1}
\plotone{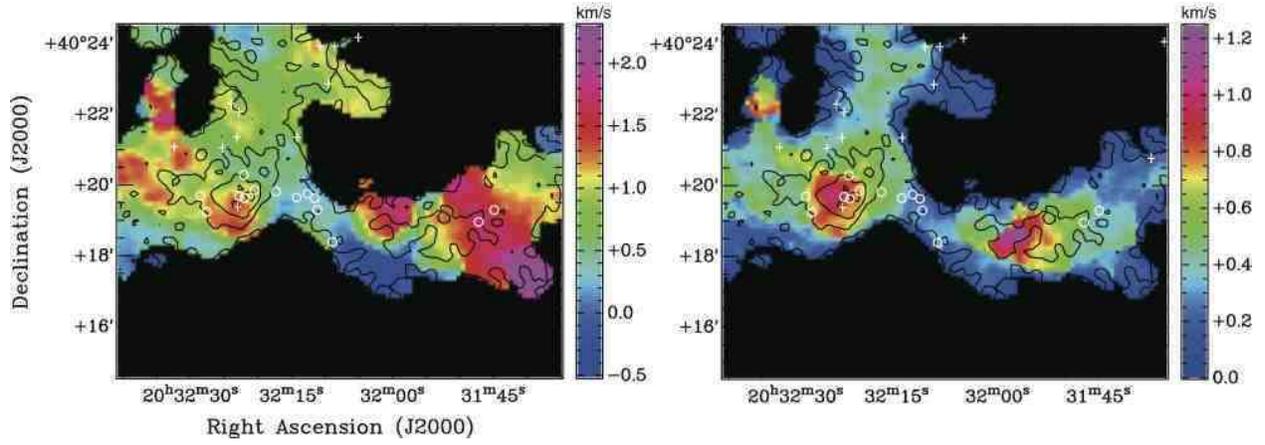}
\caption{Same as Figure \ref{mm1} for G 79.3+0.3.}
\label{mm4}
\end{figure}

\begin{figure}
\epsscale{1}
\plotone{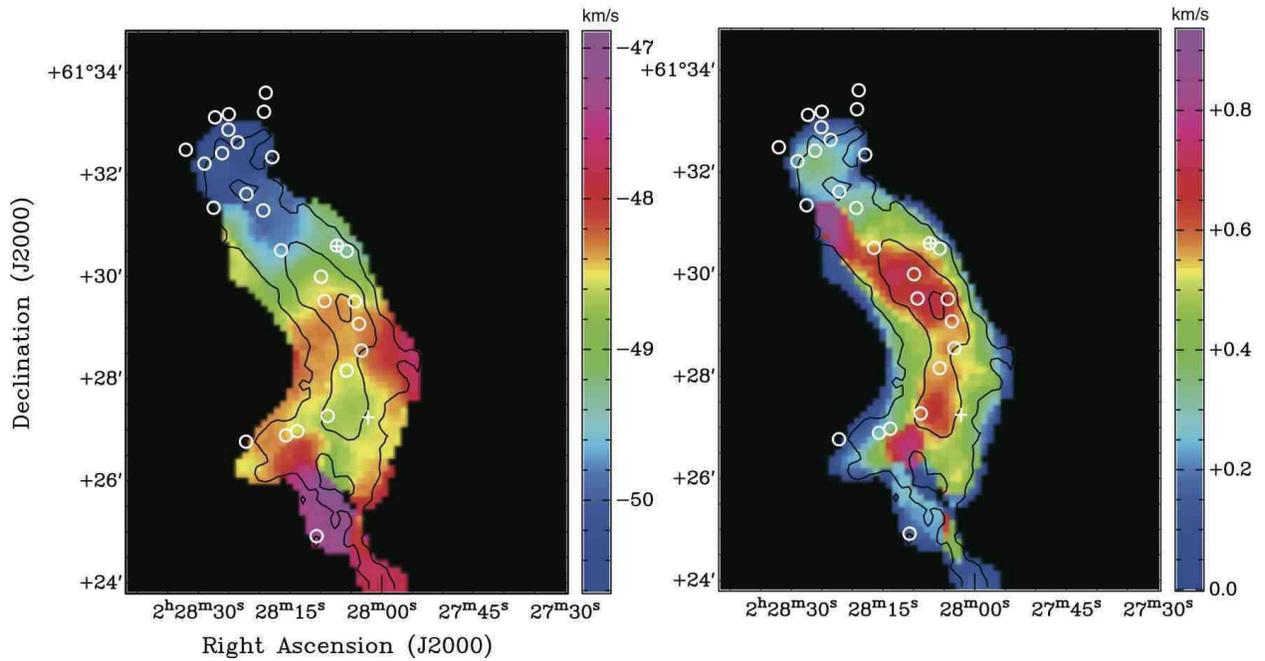}
\caption{Same as Figure \ref{mm1} for AFGL 333.}
\label{mm5}
\end{figure}

\begin{figure}
\epsscale{0.7}
\plotone{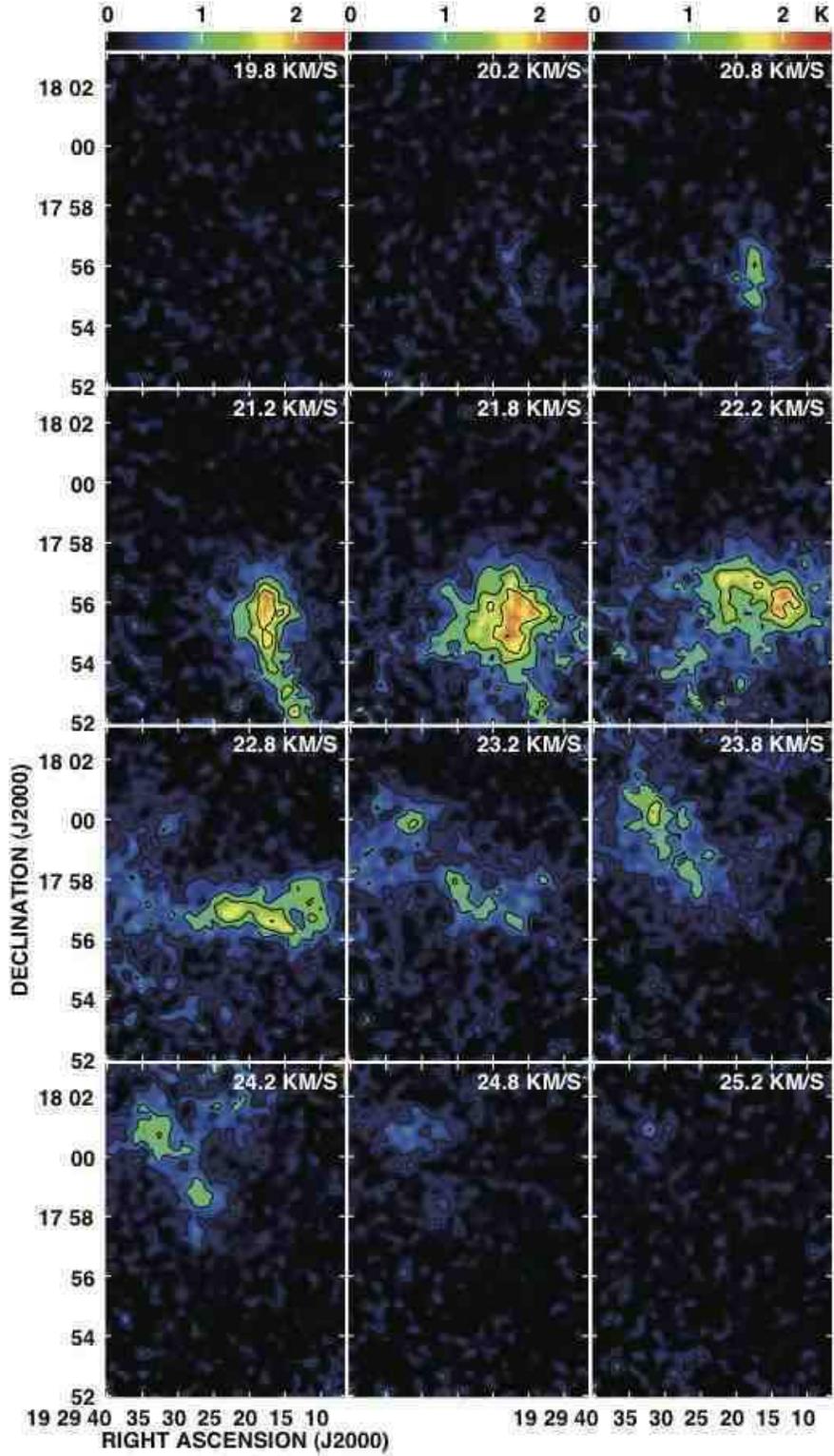}
\caption{The velocity channel maps of the $\CO$($J$=1--0) emission with contours and grayscale for MSXDC G 053.11+00.05. 
The contours with the intervals of the 3$\,$$\sigma$ levels start from 
the 3 $\sigma$ levels, where the 1$\,$$\sigma$ noise levels are 0.14 K in ${T}^{*}_{\mathrm{A}}$.
The velocity intervals are 0.4 km s$^{-1}$.}
\label{mapch1}
\end{figure}

\begin{figure}
\epsscale{0.8}
\plotone{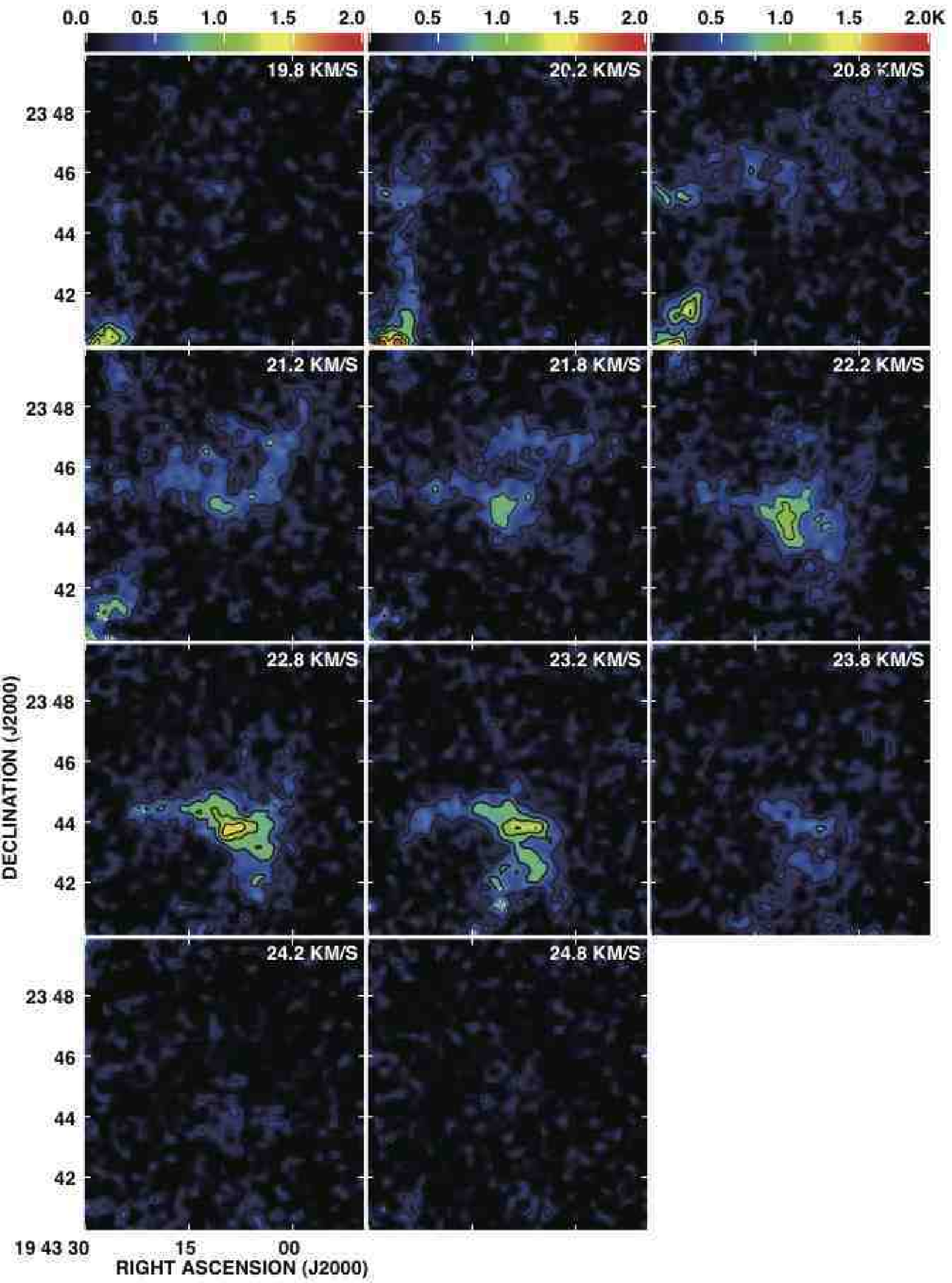}
\caption{Same as Figure \ref{mapch1} for IRDC 19410+2336.
The contours with the intervals of the 3$\,$$\sigma$ levels start from 
the 3 $\sigma$ levels, where the 1$\,$$\sigma$ noise levels are 0.11 K in ${T}^{*}_{\mathrm{A}}$.
The velocity intervals are 0.4 km s$^{-1}$.}
\label{mapch2}
\end{figure}

\begin{figure}
\epsscale{0.9}
\plotone{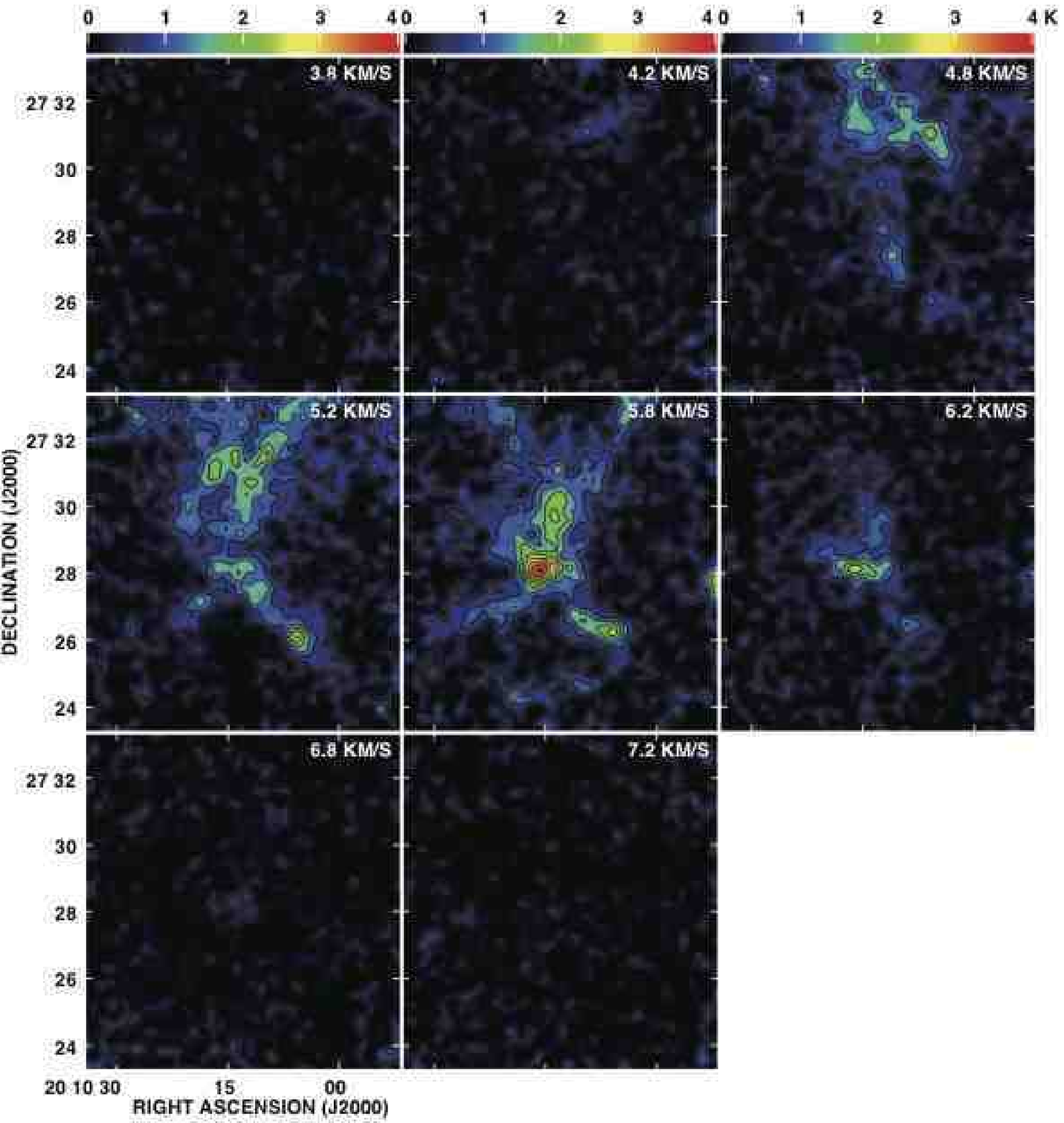}
\caption{Same as Figure \ref{mapch1} for IRDC 20081+2720.
The contours with the intervals of the 3$\,$$\sigma$ levels start from 
the 3 $\sigma$ levels, where the 1$\,$$\sigma$ noise levels are 0.14 K in ${T}^{*}_{\mathrm{A}}$.
The velocity intervals are 0.4 km s$^{-1}$.}
\label{mapch5}
\end{figure}

\begin{figure}
\epsscale{0.9}
\plotone{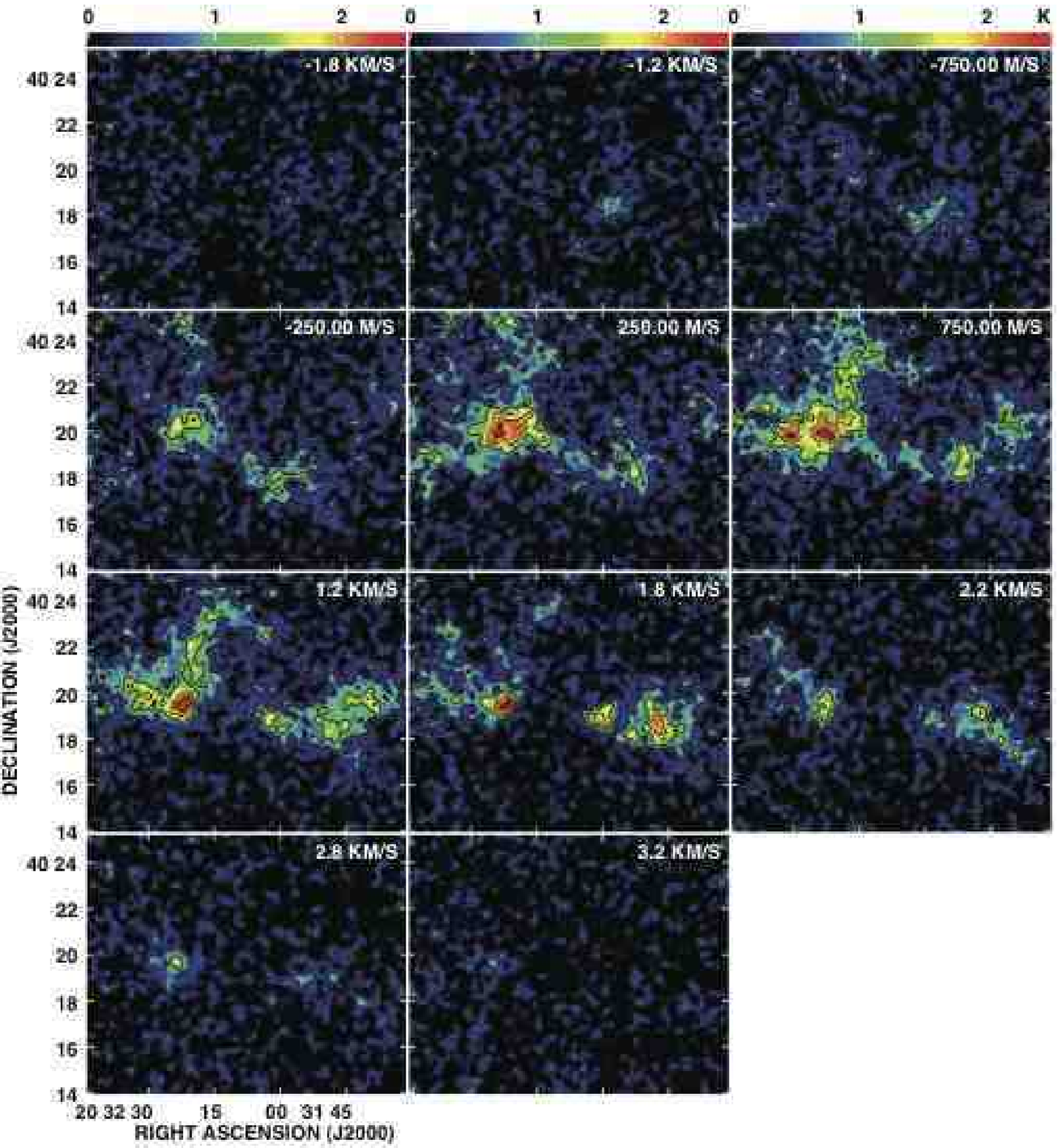}
\caption{Same as Figure \ref{mapch1} for G 79.3+0.3.
The contours with the intervals of the 3$\,$$\sigma$ levels start from 
the 3 $\sigma$ levels, where the 1$\,$$\sigma$ noise levels are 0.20 K in ${T}^{*}_{\mathrm{A}}$.
The velocity intervals are 0.4 km s$^{-1}$.}
\label{mapch3}
\end{figure}

\begin{figure}
\epsscale{0.8}
\plotone{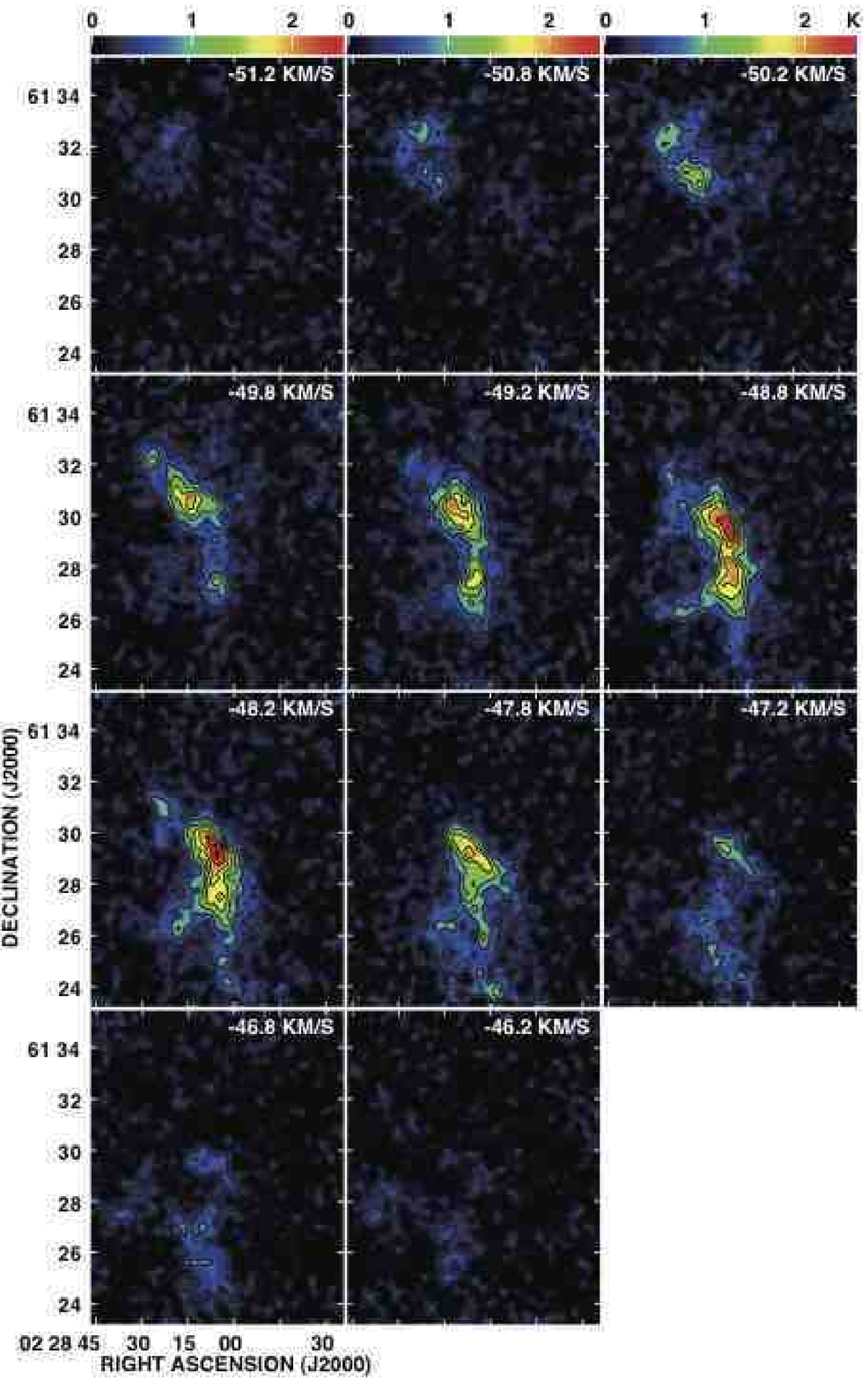}
\caption{Same as Figure \ref{mapch1} for AFGL 333.
The contours with the intervals of the 3$\,$$\sigma$ levels start from 
the 3 $\sigma$ levels, where the 1$\,$$\sigma$ noise levels are 0.12 K in ${T}^{*}_{\mathrm{A}}$.
The velocity intervals are 0.4 km s$^{-1}$.}
\label{mapch4}
\end{figure}

\begin{figure}
\epsscale{1}
\plotone{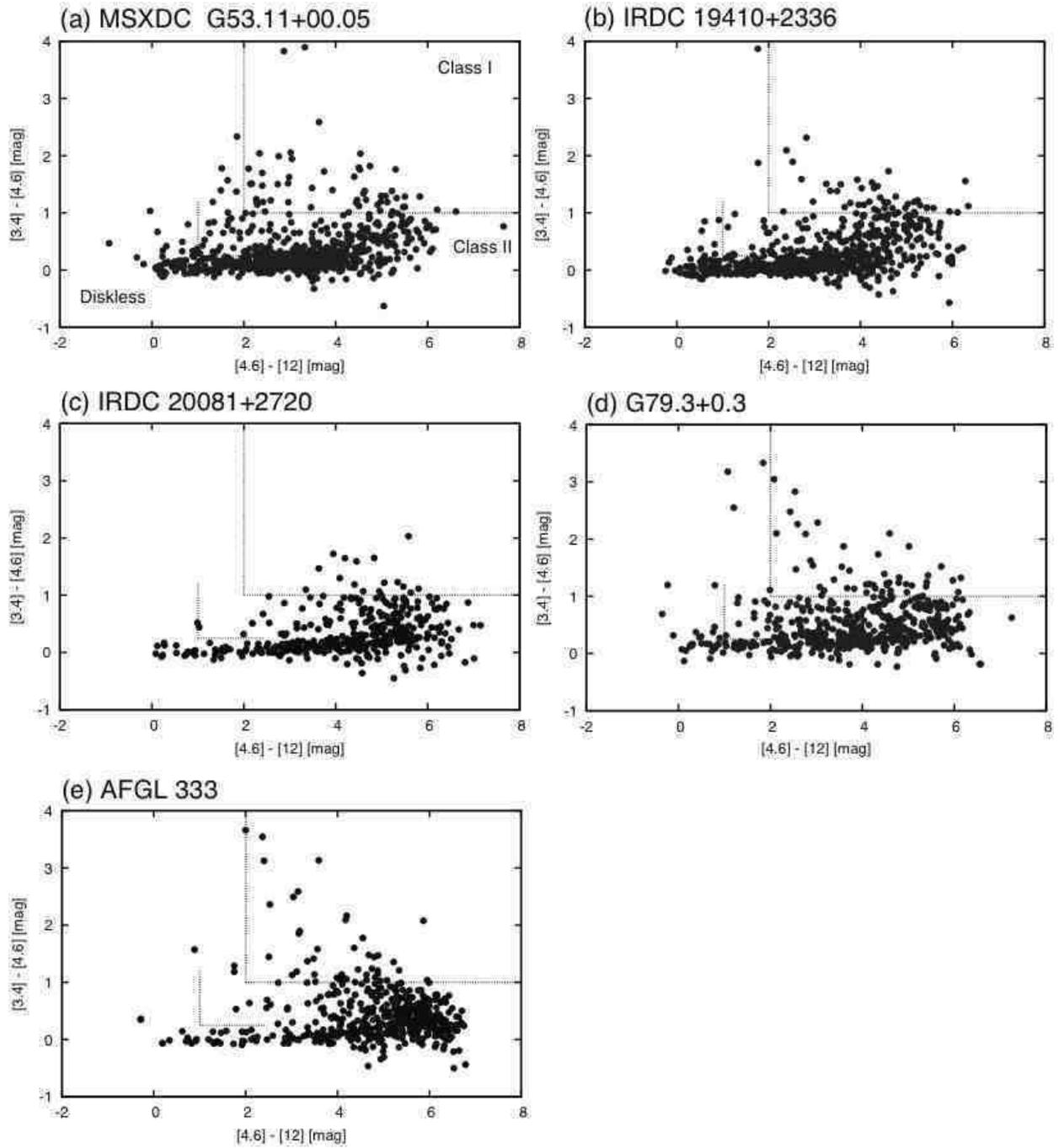}
\caption{$\it{WISE}$ 3.4 $\micron$, 4.6 $\micron$, and 12 $\micron$ color-color diagram proposed by Koenig et al. 
(2012) for 10$^{\prime}$$\times$10$^{\prime}$ areas.
Right side of dashed lines indicate the boundaries between Class I and Class I\hspace{-.1em}I candidates, and 
the left side of dashed lines indicate diskless candidates and Class I\hspace{-.1em}I candidates.}
\label{ccd_1}
\end{figure}

\begin{figure}
\epsscale{1}
\plotone{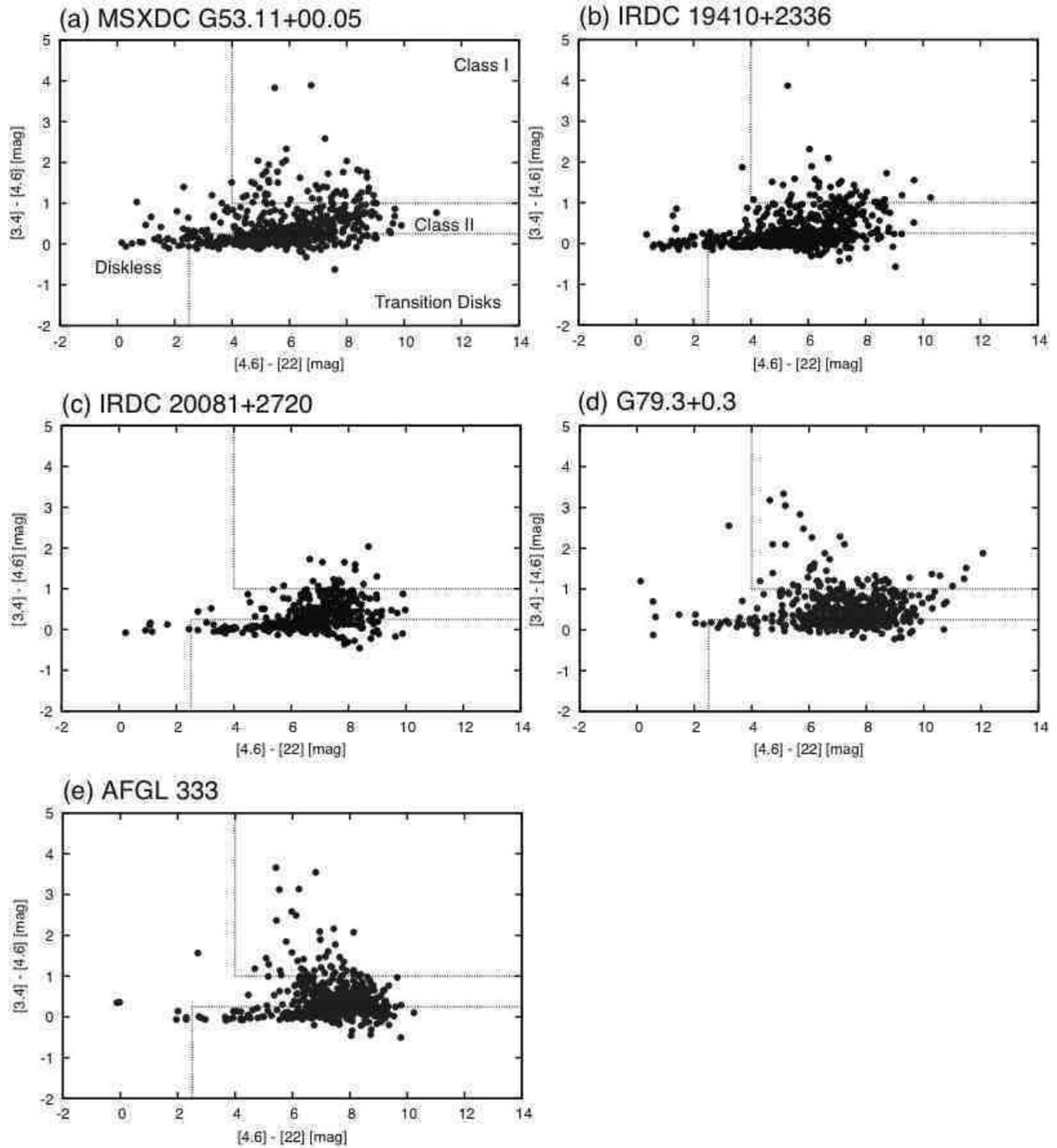}
\caption{$\it{WISE}$ 3.4 $\micron$, 4.6 $\micron$, and 22 $\micron$ color-color diagram proposed by Koenig et al. (2012) 
for 10$^{\prime}$$\times$10$^{\prime}$ areas. 
Top of dashed lines indicate the boundaries between Class I and Class I\hspace{-.1em}I candidates, and bottom lines indicate the 
boundaries between Class I and transition disk candidates.}
\label{ccd_2}
\end{figure}

\begin{figure}
\epsscale{1}
\plotone{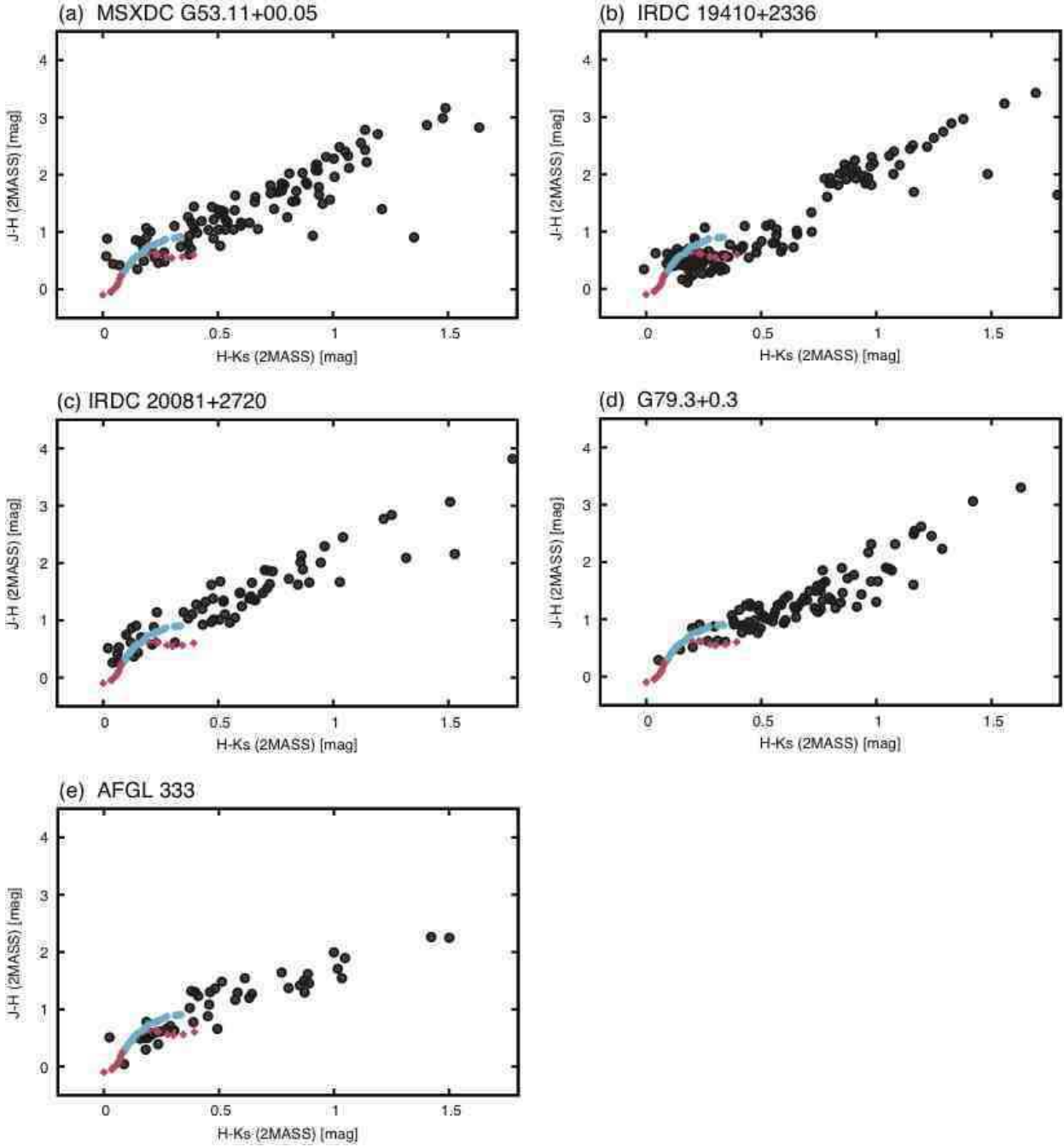}
\caption{$J$-$H$ vs. $H$-$K_{s}$ color-color diagrams of 2MASS point sources for our target objects for about 10$^{\prime}$$\times$10$^{\prime}$ areas.
The solid curves show the locus of un-reddened main sequence (pink) and giant stars (blue) in the CIT system (Bessell $\&$ Brett 1988).
The photometric system of the intrinsic colors for Dwarfs (pink) and Giants (blue) corresponded to that of 2MASS system.}
\label{ccd}
\end{figure}

\begin{figure}
\epsscale{1}
\plotone{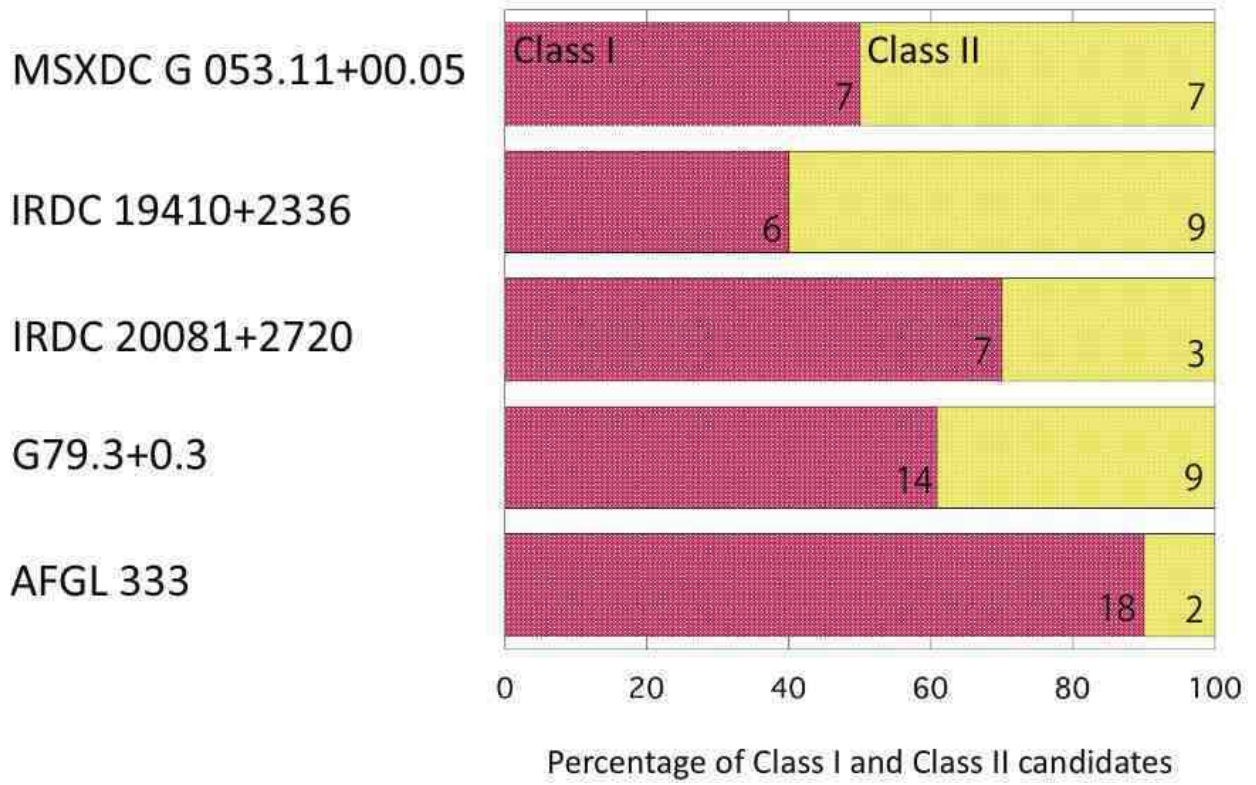}
\caption{Bar graph of the fraction of Class I (pink) and Class I\hspace{-.1em}I (yellow) candidates listed in Table \ref{sep}.
The number shows $N_{\rm{Class I}}$ and $N_{\rm{Class I\hspace{-.1em}I}}$, which are enclosed within the 
3$\,$$\sigma$ emission of $\CO$ lines contours.}
\label{histo}
\end{figure}

We can see several YSO candidates in the $\CO$ emission (Figure \ref{map5}),
whose separations are comparable to a Jeans length ($\sim$ 0.2$\,$pc).
In addition, we also see the bright infrared emission in the east of the $\CO$ emission, 
which corresponds to an H {\sc ii} region generated by a B0.5 type star \citep{hug82}.
It is located $\sim$ 1$\,$pc away from the strong $\CO$ line emission.
In the north area of the $\CO$ emission, we can see the relatively bright and diffuse 22 $\micron$ emission, 
which corresponds to a Herbig Ae/Be star \citep{sak07}.
The channel maps in Figure \ref{mapch4} clearly show that strong $\CO$ line emission is associated 
with these YSOs candidates in a range of $-$50.2 km$\,$s$^{-1}$ to $-$47.2 km$\,$s$^{-1}$.
In a velocity range from $-$48.2 km$\,$s$^{-1}$ to $-$49.2 km$\,$s$^{-1}$, the $\CO$ emission have an shell-like structure.
The YSO candidates are distributed in a shell-like fashion around the H {\sc ii} regions.
In a velocity range from $-$50.8 km$\,$s$^{-1}$ to $-$48.2 km$\,$s$^{-1}$, we can see the $\CO$ emission in northern edge have 
also an shell-like structure, which seem be compressed by the radiation or stellar wind of the Herbig Ae/Be star.

\clearpage

\clearpage

\clearpage

\clearpage

\clearpage

\clearpage

\clearpage

\clearpage

\clearpage


%
%
%
%
%
%
%
%

\begin{deluxetable}{l l l l l l l l l l l c c c c c c c c c c c c }
\tabletypesize{\tiny}
\rotate
\tablecaption{Physical parameters of the $\CO$ clumps
\label{para}}
\tablewidth{0pt}
\tablehead{
\colhead{Source Name} & \colhead{RA (J2000)} & \colhead{Dec (J2000)} & {$D$ [kpc]} &
\colhead{$\RC$[pc]} & \colhead{$\dvC$[km $\mathrm{s}^{-1}$]} & \colhead{$\MVIR$[$\MO$]} & \colhead{$\rm{N}(\mathrm{H}_{2}$)[$\times$ 10$^{22}$ $\mathrm{cm}^{-2}$]} 
& \colhead{$\MCLU$[$\MO$]} & \colhead{$\rm{T}_{\rm{ex}}$ [$K$]} & \colhead{H {\sc ii} region}}
\startdata
 MSXDC G53.11+00.05 C1 & 19h29m18s & 17d56m33s & 1.8 & 1.2$\pm$0.4 & 1.9$\pm$0.3 & 920$\pm$560 & 3.4$\pm$0.7 & 3200$\pm$1400 & 18 & No \\
 MSXDC G53.11+00.05 C2 & 19h29m37s & 18d02m00s & 1.9 & 0.7$\pm$0.2 & 2.0$\pm$0.3 & 610$\pm$360  & 2.4$\pm$0.6 & 740$\pm$370 & 15 & No \\
 IRDC 19410+2336 & 19h43m10s & 23d45m04s & 2.1 & 1.7$\pm$0.5 & 3.3$\pm$0.3 & 4000$\pm$2000  & 1.7$\pm$0.5 & 2900$\pm$1200 & 15 & No  \\
 IRDC 20081+2720 & 20h10m13s & 27d28m18s & 0.7 & 0.6$\pm$0.1 & 1.4$\pm$0.3 & 230$\pm$140  & 2.6$\pm$0.4 & 470$\pm$150 & 15$\tablenotemark{a}$ & Yes \\
 G79.3+0.3 C1 & 20h32m23s & 40d20m00s & 0.8 & 0.7$\pm$0.1 & 2.2$\pm$0.3 & 650$\pm$310 & 4.2$\pm$0.5 & 1000$\pm$340 & 15$\tablenotemark{a}$ & Yes \\
 G79.3+0.3 C2 & 20h32m06s & 40d18m00s & 0.8 & 0.5$\pm$0.1 & 2.8$\pm$0.3 & 860$\pm$360 & 3.9$\pm$0.5 & 630$\pm$230 & 15$\tablenotemark{a}$ & Yes \\
 AFGL 333 & 02h28m19s & 61d29m40s & 2.0 & 1.4$\pm$0.4 & 3.0$\pm$0.3 & 2500$\pm$1300 & 3.7$\pm$0.4 & 4200$\pm$1800 & 15 & Yes \\
\hline
\enddata
{\small
\tablecomments{This table shows the name, center coordinates, distance, mass of the clusters,
$\RC$: radius, $\dvC$: velocity width, $N(\rm{H}_{2})$: mean $\rm{H}_{2}$ column density, 
$\MCLU$: clump mass, $T_{\rm{ex}}$: excitation temperature derived by kinetic temperature of NH$_{3}$ of the clumps.}}
\tablenotetext{a}{Because we could not estimate the excitation temperature for low signal-to-noise ratio of the NH$_{3}$ data,
we adopted 15$\,\rm K$ which is the typical temperature in IRDCs regions.}
\end{deluxetable}

\begin{deluxetable}{l l l l l l c c c c c}
\tabletypesize{\scriptsize}
\tablecaption{Kinetic energies of $\CO$ clumps
\label{energy}}
\tablewidth{0pt}
\tablehead{
\colhead{Source Name} & \colhead{$\lambda_{\rm{J}}$ [pc]} & \colhead{$\tau_{\rm{cross}}$ [yr]} & \colhead{$\rm{E}_{\rm{kin}}$ [erg]} &
\colhead{$\rm{E}_{\rm{kin}}/\tau_{\rm{cross}}$ [erg s$^{-1}$]}}
\startdata
 MSXDC G53.11+00.05 C1 & 0.21 & 3$\times$10$^{6}$ & 2.1$\times$10$^{46}$ & 2.2$\times$10$^{32}$ \\
 MSXDC G53.11+00.05 C2 & 0.17 & 2$\times$10$^{6}$ & 5.6$\times$10$^{45}$ & 1.1$\times$10$^{32}$ \\
 IRDC 19410+2336 & 0.34 & 2$\times$10$^{6}$ & 5.9$\times$10$^{46}$ & 7.9$\times$10$^{32}$ \\
 IRDC 20081+2720 & 0.16 & 2$\times$10$^{6}$ & 1.7$\times$10$^{45}$ & 2.9$\times$10$^{31}$ \\
 G79.3+0.3 C1 & 0.13 & 1$\times$10$^{6}$ & 8.6$\times$10$^{45}$ & 2.0$\times$10$^{32}$ \\
 G79.3+0.3 C2 & 0.12 & 9$\times$10$^{5}$ & 8.7$\times$10$^{45}$ & 3.1$\times$10$^{32}$ \\
 AFGL 333 & 0.20 & 2$\times$10$^{6}$ & 6.6$\times$10$^{46}$ & 9.8$\times$10$^{32}$ \\
\hline
\enddata
{\small
\tablecomments{This table shows the source name, the effective Jeans length of the clumps, 
the effective crossing time, the internal kinetic energy of the clumps, and the turbulence dissipation rate of the clumps.}}
\end{deluxetable}

\begin{deluxetable}{ l l l l l c c c c}
\tabletypesize{\scriptsize}
\tablecaption{Information of YSO candidates within enclosed 3$\,$$\sigma$ contours (except within H {\sc ii} region)
\label{sep}}
\tablewidth{0pt}
\tablehead{
\colhead{Source Name} & 
\colhead{$\rm{N}_{\rm{Class I}}$} & \colhead{$\rm{N}_{\rm{Class I\hspace{-.1em}I}}$} &  \colhead{$\rm{N}_{\rm{2MASS}}$} & 
\colhead{$\rm{N}_{\rm{Class I}}$/($\rm{N}_{\rm{Class I\hspace{-.1em}I}}$+$\rm{N}_{\rm{Class I}}$) [$\%$]}}
\startdata
 MSXDC G53.11+00.05 C1 & 4 (6) & 3 (7) & 3 & 57  \\
 MSXDC G53.11+00.05 C2 &  3 (4) & 4 (7) & 0 & 43 \\
 MSXDC G53.11+00.05 (total) &  7 (10) & 7 (14) & 3 & 50 \\
 IRDC 19410+2336 &  6 & 9 (11) & 3 & 40 \\ 
 IRDC 20081+2720 &  7 & 3 & 0 & 70 \\
 G79.3+0.3 C1 & 11 & 9 & 0 & 55 \\
 G79.3+0.3 C2 &  3 & 0 & 0 & 100 \\
 G79.3+0.3 (total) & 14 & 9 & 0 & 61 \\
 AFGL 333 & 18 (21) & 2 & 0 & 90 \\
\hline
\enddata
{\small
\tablecomments{
${\rm{N}_{\rm{Class I}}}$: the number of Class I candidates within the clump, 
${\rm{N}_{\rm{Class I\hspace{-.1em}I }}}$: the number of Class I\hspace{-.1em}I  candidates within the clump.
The numbers in parentheses show total numbers of YSOs candidates, which are distributed enclosed 3$\,$$\sigma$ contours and
distributed adjacent 3$\,$$\sigma$ contours.
${\rm{N}_{\rm{2MASS}}}$: the number of 2MASS YSO candidates within the clump, and
${\rm{N}_{\rm{Class I}}}$/(${\rm{N}_{\rm{Class I}}}$+${\rm{N}_{\rm{Class I\hspace{-.1em}I}}}$): the fraction of Class I to the YSO candidates.}}
\end{deluxetable}

\begin{deluxetable}{l l l c c}
\tabletypesize{\scriptsize}
\tablecaption{Score sheet of different features between CWHRs and CWOHRs
\label{score}}
\tablewidth{0pt}
\tablehead{
\colhead{} & \colhead{CWHRs (H {\sc ii} region)} & \colhead{CWOHRs (Non-H {\sc ii} region)}}
\startdata
 Source Name & IRDC 20081+2720, G79.3+0.3, AFGL 333 & MSXDC G53.11+00.05, IRDC 19410+2336 \\
 Spatial distribution & filamentary, shell-like & spherical  \\
 Velocity gradient & present ($\sim$ 2 km$\,$s$^{-1}$pc$^{-1}$)  & absent \\
 Velocity dispersion & large ($\sim$ 1$\,$km$\,$s$^{-1}$)$\tablenotemark{a}$along shell-like structure & large ($\sim$ 1$\,$km$\,$s$^{-1}$)$\tablenotemark{a}$at central region  \\
 Age of YSO association & $\leq$ 0.5$\,$Myr & $\sim$ 1$\,$Myr \\
\hline
\enddata
{\small
\tablecomments{This table shows a summary of the differences between CWHRs and CWOHRs;
spatial distributions, velocity structures, and age of YSO association for our target objects.}}
\tablenotetext{a}{This value is derived from the 2nd moment maps in Figure \ref{mm1} to \ref{mm5}.}
\end{deluxetable}


\begin{thebibliography}{}

\bibitem[Abbott(1982)]{abb82} Abbott, D.~C.\ 1982, \apj, 263, 723 

\bibitem[Allen et al.(2007)]{all07} Allen, L., Megeath, S.~T., Gutermuth, R., et al.\ 2007, Protostars and Planets V, 361 

\bibitem[Bessell \& Brett(1988)]{bes88} Bessell, M.~S., \& Brett, J.~M.\ 1988, \pasp, 100, 1134 
\bibitem[Beuther et al.(2003)]{beu03} Beuther, H., Schilke, P., \& Stanke, T.\ 2003, \aap, 408, 601 
\bibitem[Beuther et al.(2002)]{beu02a} Beuther, H., Schilke, P., Menten, K.~M., Motte, F., Sridharan, T.~K., \& Wyrowski, F.\ 2002, \apj, 566, 945 
\bibitem[Bik et al.(2012)]{bik12} Bik, A., Henning, T., Stolte, A., et al.\ 2012, \apj, 744, 87 
\bibitem[Bonnell et al.(2003)]{bon03} Bonnell, I.~A., Bate, M.~R., \& Vine, S.~G.\ 2003, \mnras, 343, 413 
\bibitem[Brand et al.(2011)]{bra11} Brand, J., Massi, F., Zavagno, A., Deharveng, L., \& Lefloch, B.\ 2011, \aap, 527, A62 
\bibitem[Carey et al.(1998)]{car98} Carey, S.~J., Clark, F.~O., Egan, M.~P., Price, S.~D., Shipman, R.~F., \& Kuchar, T.~A.\ 1998, \apj, 508, 721 
\bibitem[Carey et al.(2000)]{car00b} Carey, S.~J., Feldman, P.~A., Redman, R.~O., Egan, M.~P., MacLeod, J.~M., \& Price, S.~D.\ 2000, \apjl, 543, L157 
\bibitem[Carpenter(2000)]{car00} Carpenter, J.~M.\ 2000, \aj, 120, 3139 
\bibitem[Churchwell et al.(2007)]{chu07} Churchwell, E., Watson, D.~F., Povich, M.~S., et al.\ 2007, \apj, 670, 428 
\bibitem[Churchwell et al.(2006)]{chu06} Churchwell, E., et al.\ 2006, \apj, 649, 759 
\bibitem[Cohen et al.(1981)]{coh81} Cohen, J.~G., Persson, S.~E., Elias, J.~H., \& Frogel, J.~A.\ 1981, \apj, 249, 481 
\bibitem[Deharveng et al.(2010)]{deh10} Deharveng, L., et al.\ 2010, \aap, 523, A6 
\bibitem[Deharveng et al.(2009)]{deh09} Deharveng, L., Zavagno, A., Schuller, F., Caplan, J., Pomar{\`e}s, M., \& De Breuck, C.\ 2009, \aap, 496, 177 
\bibitem[Deharveng et al.(2005)]{deh05} Deharveng, L., Zavagno, A., \& Caplan, J.\ 2005, \aap, 433, 565 
\bibitem[Egan et al.(1998)]{ega98} Egan, M.~P., Shipman, R.~F., Price, S.~D., Carey, S.~J., Clark, F.~O., \& Cohen, M.\ 1998, \apjl, 494, L199 
\bibitem[Elmegreen(2000)]{elm00} Elmegreen, B.~G.\ 2000, \apj, 530, 277 
\bibitem[Elmegreen \& Lada(1977)]{elm77} Elmegreen, B.~G., \& Lada, C.~J.\ 1977, \apj, 214, 725 
\bibitem[Evans et al.(2009)]{eva09} Evans, N.~J., II, Dunham,  M.~M., J{\o}rgensen, J.~K., et al.\ 2009, \apjs, 181, 321 
\bibitem[Frerking et al.(1982)]{fre82} Frerking, M.~A., Langer, W.~D., \& Wilson, R.~W.\ 1982, \apj, 262, 590 
\bibitem[Fuente et al.(1998)]{fue98} Fuente, A., Martin-Pintado, J., Bachiller, R., Neri, R., \& Palla, F.\ 1998, \aap, 334, 253 
\bibitem[Fuente et al.(2002)]{fue02} Fuente, A., Mart{\i}n-Pintado, J., Bachiller, R., Rodr{\i}guez-Franco, A., \& Palla, F.\ 2002, \aap, 387, 977 
\bibitem[Fukuda \& Hanawa(2000)]{fuk00} Fukuda, N., \& Hanawa, T.\ 2000, \apj, 533, 911 

\bibitem[Gutermuth et al.(2008)]{gut08} Gutermuth, R.~A., Bourke, T.~L., Allen, L.~E., et al.\ 2008, \apjl, 673, L151 
\bibitem[Gutermuth et al.(2009)]{gut09} Gutermuth, R.~A., Megeath, S.~T., Myers, P.~C., et al.\ 2009, \apjs, 184, 18 

\bibitem[Hartmann(2002)]{har02} Hartmann, L.\ 2002, \apj, 578, 914 

\bibitem[Higuchi et al.(2009)]{hig09} Higuchi, A.~E., Kurono, Y., Saito, M., \& Kawabe, R.\ 2009, \apj, 705, 468 
\bibitem[Higuchi et al.(2010)]{hig10} Higuchi, A.~E., Kurono, Y., Saito, M., \& Kawabe, R.\ 2010, \apj, 719, 1813 
\bibitem[Hosokawa \& Inutsuka(2006)]{hos06} Hosokawa, T., \& Inutsuka, S.-i.\ 2006, \apj, 646, 240 
\bibitem[Ho \& Townes (1983)]{ho83} Ho, P.~T.~P., \& Townes, C.~H.\ 1983, \araa, 21, 239 
\bibitem[Hofner et al.(2000)]{hof00} Hofner, P., Wyrowski, F., Walmsley, C.~M., \& Churchwell, E.\ 2000, \apj, 536, 393 
\bibitem[Hughes \& Viner(1982)]{hug82} Hughes, V.~A., \& Viner, M.~R.\ 1982, \aj, 87, 685 
\bibitem[Ikeda et al.(2007)]{ike07} Ikeda, N., Sunada, K., \& Kitamura, Y.\ 2007, \apj, 665, 1194 
\bibitem[Inutsuka \& Miyama(1992)]{inu92} Inutsuka, S.-I., \& Miyama, S.~M.\ 1992, \apj, 388, 392 
\bibitem[Jarrett et al.(2011)]{jar11} Jarrett, T.~H., Cohen, M., Masci, F., et al.\ 2011, \apj, 735, 112 
\bibitem[Kauffmann \& Pillai(2010)]{kau10} Kauffmann, J., \& Pillai, T.\ 2010, \apjl, 723, L7 
\bibitem[Kessel-Deynet \& Burkert(2003)]{kes03} Kessel-Deynet, O., \& Burkert, A.\ 2003, \mnras, 338, 545 
\bibitem[Knee \& Sandell(2000)]{kne00} Knee, L.~B.~G., \& Sandell, G.\ 2000, \aap, 361, 671 
\bibitem[Koenig et al.(2012)]{koe12} Koenig, X.~P., Leisawitz, D.~T., Benford, D.~J., et al.\ 2012, \apj, 744, 130 
\bibitem[Lada \& Lada(2003)]{lad03} Lada, C.~J., \& Lada, E.~A.\ 2003, \araa, 41, 57 
\bibitem[Lada(2010)]{lad10} Lada, C.~J.\ 2010, Royal Society of London Philosophical Transactions Series A, 368, 713 
\bibitem[Lefloch \& Lazareff(1994)]{lef94} Lefloch, B., \& Lazareff, B.\ 1994, \aap, 289, 559 
\bibitem[L{\'o}pez-Sepulcre et al.(2010)]{lop10} L{\'o}pez-Sepulcre, A., Cesaroni, R., \& Walmsley, C.~M.\ 2010, \aap, 517, A66 
\bibitem[Mart{\'{\i}}n-Hern{\'a}ndez et al.(2008)]{mar08} Mart{\'{\i}}n-Hern{\'a}ndez, N.~L., Bik, A., Puga, E., N{\"u}rnberger, D.~E.~A., \& Bronfman, L.\ 2008, \aap, 489, 229 
\bibitem[Mauersberger \& Henkel(1993)]{mau93} Mauersberger, R., \& Henkel, C.\ 1993, Reviews in Modern Astronomy, 6, 69 
\bibitem[Niwa et al.(2009)]{niw09} Niwa, T., Tachihara, K., Itoh, Y., Oasa, Y., Sunada, K., Sugitani, K., \& Mukai, T.\ 2009, \aap, 500, 1119 
\bibitem[Nakamura et al.(2011)]{nak11} Nakamura, F., Sugitani, K., Shimajiri, Y., et al.\ 2011, \apj, 737, 56 
\bibitem[Oka et al.(2001)]{oka01} Oka, T., Yamamoto, S., Iwata, M., et al.\ 2001, \apj, 558, 176 
\bibitem[Odenwald et al.(1990)]{ode90} Odenwald, S.~F., Campbell, M.~F., Shivanandan, K., et al.\ 1990, \aj, 99, 288 
\bibitem[Peretto \& Fuller(2009)]{per09} Peretto, N., \& Fuller, G.~A.\ 2009, \aap, 505, 405 

\bibitem[Pineda et al.(2011)]{pie11} Pineda, J.~E., Goodman, A.~A., Arce, H.~G., et al.\ 2011, \apjl, 739, L2 

\bibitem[Pomar{\`e}s et al.(2009)]{pom09} Pomar{\`e}s, M., Zavagno, A., Deharveng, L., et al.\ 2009, \aap, 494, 987 
\bibitem[Puga et al.(2009)]{pug09} Puga, E., Hony, S., Neiner, C., et al.\ 2009, \aap, 503, 107 
\bibitem[Plume et al.(1997)]{plu97} Plume, R., Jaffe, D.~T., Evans, N.~J., II, Martin-Pintado, J., 
\& Gomez-Gonzalez, J.\ 1997, \apj, 476, 730 
\bibitem[Rathborne et al.(2010)]{rat10} Rathborne, J.~M., Jackson, J.~M., Chambers, E.~T., Stojimirovic, I., Simon, R., Shipman, R., 
\& Frieswijk, W.\ 2010, \apj, 715, 310 
\bibitem[Rathborne et al.(2006)]{rat06} Rathborne, J.~M., Jackson, J.~M., \& Simon, R.\ 2006, \apj, 641, 389 
\bibitem[Ragan et al.(2012)]{rag12} Ragan, S.~E., Heitsch, F., Bergin, E.~A., \& Wilner, D.\ 2012, \apj, 746, 174 
\bibitem[Redman et al.(2003)]{red03} Redman, R.~O., Feldman, P.~A., Wyrowski, F., C{\^o}t{\'e}, S., Carey, S.~J., 
\& Egan, M.~P.\ 2003, \apj, 586, 1127 
\bibitem[Rizzo et al.(2008)]{riz08} Rizzo, J.~R., Jim{\'e}nez-Esteban, F.~M., \& Ortiz, E.\ 2008, \apj, 681, 355
\bibitem[Ridge et al.(2003)]{rid03} Ridge, N.~A., Wilson, T.~L., Megeath, S.~T., Allen, L.~E., \& Myers, P.~C.\ 2003, \aj, 126, 286 
\bibitem[Sakai et al.(2007)]{sak07} Sakai, T., Oka, T., \& Yamamoto, S.\ 2007, \apj, 662, 1043 
\bibitem[Sawada et al.(2008)]{saw08} Sawada, T., et al.\ 2008, \pasj, 60, 445 



\bibitem[Simon et al.(2006)]{sim06} Simon, R., Jackson, J.~M., Rathborne, J.~M., \& Chambers, E.~T.\ 2006, \apj, 639, 227 
\bibitem[Sorai et al.(2000)]{sor00} Sorai, K., Sunada, K., Okumura, S.~K., Tetsuro, I., Tanaka, A., Natori, K., \& Onuki, H.\ 2000, \procspie, 4015, 86 
\bibitem[Sridharan et al.(2005)]{sri05} Sridharan, T.~K., Beuther, H., Saito, M., Wyrowski, F., \& Schilke, P.\ 2005, \apjl, 634, L57 

\bibitem[Stern et al.(2005)]{ste05} Stern, D., Eisenhardt, P., Gorjian, V., et al.\ 2005, \apj, 631, 163 

\bibitem[Stahler \& Palla(2005)]{sta05} Stahler, S.~W., \& Palla, F.\ 2005, The Formation of Stars, by Steven W.~Stahler, Francesco Palla, pp.~865.~ISBN 3-527-40559-3.~Wiley-VCH


\bibitem[Sunada et al.(2000)]{sun00} Sunada, K., Yamaguchi, C., Nakai, N., Sorai, K., Okumura, S.~K., \& Ukita, N.\ 2000, \procspie, 4015, 237 

\bibitem[Teixeira et al.(2006)]{tei06} Teixeira, P.~S., Lada, C.~J., Young, E.~T., et al.\ 2006, \apjl, 636, L45 

\bibitem[Teixeira et al.(2007)]{tei07} Teixeira, P.~S., Zapata, L.~A., \& Lada, C.~J.\ 2007, \apjl, 667, L179 


\bibitem[Watson et al.(2008)]{wat08} Watson, C., Povich, M.~S., Churchwell, E.~B., et al.\ 2008, \apj, 681, 1341 
\bibitem[Whitworth et al.(1994)]{whi94} Whitworth, A.~P., Bhattal, A.~S., Chapman, S.~J., Disney, M.~J., \& Turner, J.~A.\ 1994, \mnras, 268, 291 
\bibitem[Wright et al.(2010)]{wri10} Wright, E.~L., Eisenhardt, P.~R.~M., Mainzer, A.~K., et al.\ 2010, \aj, 140, 1868 
\bibitem[Yamaguchi et al.(2000)]{yam00} Yamaguchi, C., Sunada, K., Iizuka, Y., Iwashita, H., \& Noguchi, T.\ 2000, \procspie, 4015, 614 
\bibitem[Zavagno et al.(2006)]{zav06} Zavagno, A., Deharveng, L., Comer{\'o}n, F., Brand, J., Massi, F., Caplan, J., \& Russeil, D.\ 2006, \aap, 446, 171 
\bibitem[Zavagno et al.(2007)]{zav07} Zavagno, A., Pomar{\`e}s, M., Deharveng, L., et al.\ 2007, \aap, 472, 835 
\bibitem[Zavagno et al.(2010)]{zav10} Zavagno, A., Russeil, D., Motte, F., et al.\ 2010, \aap, 518, L81 
\bibitem[Zinnecker \& Yorke(2007)]{zin07} Zinnecker, H., \& Yorke, H.~W.\ 2007, \araa, 45, 481 


\end{thebibliography}
\end{document}